\documentclass[12pt]{amsart}
\usepackage{amssymb}
\usepackage{stmaryrd}
\usepackage{isolatin1}    
\usepackage{graphics}
\usepackage{color}
\usepackage[all]{xy}

\newtheorem{thm}{Theorem}[section]
\newtheorem{lem}[thm]{Lemma}
\newtheorem{cor}[thm]{Corollary}
\newtheorem{prop}[thm]{Proposition}
\newtheorem{ex}[thm]{Example}

\newtheorem*{prob*}{Open problem}

\newtheorem{conm}[thm]{Milnor's Question}

\theoremstyle{definition}

\newtheorem{defi}[thm]{Definition}

\theoremstyle{remark}

\newtheorem{rem}[thm]{Remark}
\newtheorem*{rem*}{Remark}

\DeclareMathOperator{\id}{id}
\DeclareMathOperator{\s}{span}
\DeclareMathOperator{\rad}{rad}
\DeclareMathOperator{\sol}{sol}
\DeclareMathOperator{\nil}{nil}
\DeclareMathOperator{\Hom}{Hom}
\DeclareMathOperator{\Aff}{Aff}

\newcommand{\kringel}{\mathbin{\raise1pt\hbox{$\scriptstyle\circ$}}} 
\newcommand{\pkt}{\mathbin{\raise0pt\hbox{$\scriptstyle\bullet$}}}

\newcommand{\C}{\mathbb{C}}

\newcommand{\I}{\mathbf{1}}

\newcommand{\N}{\mathbb{N}}

\newcommand{\R}{\mathbb{R}}

\newcommand{\Z}{\mathbb{Z}}

\newcommand{\tr}{\mathop{\rm tr}}

\newcommand{\End}{\mathop{\rm End}}
\newcommand{\Der}{\mathop{\rm Der}}

\newcommand{\La}{\mathfrak{a}}

\newcommand{\Lf}{\mathfrak{f}}
\newcommand{\Lg}{\mathfrak{g}}
\newcommand{\Lh}{\mathfrak{h}}

\newcommand{\Ll}{\mathfrak{l}}

\newcommand{\Lr}{\mathfrak{r}}
\newcommand{\Ls}{\mathfrak{s}}
\newcommand{\Lt}{\mathfrak{t}}
\newcommand{\Lz}{\mathfrak{z}}

\newcommand{\LX}{\mathfrak{X}}

\newcommand{\CC}{\mathcal{C}}
\newcommand{\CD}{\mathcal{D}}
\newcommand{\CE}{\mathcal{E}}
\newcommand{\CF}{\mathcal{F}}
\newcommand{\CH}{\mathcal{H}}

\newcommand{\CL}{\mathcal{L}}

\newcommand{\CP}{\mathcal{P}}
\newcommand{\CR}{\mathcal{R}}
\newcommand{\CS}{\mathcal{S}}
\newcommand{\CT}{\mathcal{T}}

\newcommand{\abs}[1]{\lvert#1\rvert}

\newcommand{\al}{\alpha}
\newcommand{\be}{\beta}
\newcommand{\ga}{\gamma}
\newcommand{\de}{\delta}
\newcommand{\ep}{\varepsilon}

\newcommand{\la}{\lambda}
\newcommand{\om}{\omega}

\newcommand{\Ga}{\Gamma}

\newcommand{\ra}{\rightarrow}  

\renewcommand{\phi}{\varphi}

\begin{document}


\title[LSAs in geometry and physics]{Left-symmetric algebras, or pre-Lie algebras in geometry and physics}
\author[D. Burde]{Dietrich Burde}

\address{Fakult\"at f\"ur Mathematik\\
Universit\"at Wien\\
  Nordbergstrasse 15\\
  1090 Wien}

\date{\today}
\email{dietrich.burde@univie.ac.at}


\begin{abstract}
In this survey article we discuss the origin, theory and applications of left-symmetric algebras
(LSAs in short) in geometry in physics. Recently Connes, Kreimer and Kontsevich have introduced
LSAs in mathematical physics (QFT and renormalization theory), where the name pre-Lie
algebras is used quite often. Already Cayley wrote about such algebras more than
hundred years ago. Indeed, LSAs arise in many different areas of mathematics and physics.
We attempt to give a survey of the fields where LSAs play an important role.
Furthermore we study the algebraic theory of LSAs such as structure theory, 
radical theory, cohomology theory and the classification of simple LSAs.
We also discuss applications to faithful Lie algebra representations.
\end{abstract}

\maketitle

\tableofcontents

\section{Introduction}

Left-symmetric algebras, or LSAs in short, arise in many areas of mathematics
and physics. They have already been introduced by A. Cayley in $1896$, in the
context of rooted tree algebras, see \cite{CAY}. Then they were forgotten for
a long time until Vinberg \cite{VIN} in $1960$ (in the original russian version)
and Koszul \cite{KOS} in $1961$ introduced them in the context of 
convex homogeneous cones and affinely flat manifolds. From this time on many
articles related to LSAs, from quite different research areas, have been published. 
As a consequence, perhaps, LSAs are known under many different names. 
LSAs are also called Vinberg algebras, Koszul algebras or quasi-associative
algebras. Right-symmetric algebras, or RSAs, are also called Gerstenhaber algebras, or 
pre-Lie algebras \cite{CHL}. 
The aim of the first section is to give a survey on the main topics
involving LSAs and to describe the role of LSAs therein. The importance
of LSAs for the subject may be quite different. In the problems comming
from differential geometry LSAs have been introduced in order to reformulate
the geometric problem in terms of algebra. In this case the original problem
is equivalent to a certain problem on LSAs. For other problems
a combinatorically defined product turns out to be left- or right-symmetric,
but the importance of this structure is not always obvious. \\
There exist also many attempts to provide a structure theory for finite-dimen\-sional
LSAs over the real or complex numbers. We will describe known results on
the algebraic theory of LSAs and its applications in the second section.\\[0.2cm]
We start with some basic definitions which we already need for the first section.  
Let $(A,\cdot)$ be an algebra over $K$, not necessarily associative and 
not necessarily finite-dimensional. The
associator $(x,y,z)$ of three elements $x,y,z\in A$ is defined by
\begin{align*}
(x,y,z) & = (x\cdot y) \cdot z- x \cdot (y \cdot z).
\end{align*}

\begin{defi}
An algebra $(A,\cdot)$ over $K$ with a bilinear product $(x,y) \mapsto x\cdot y$
is called {\it LSA}, if the product is left-symmetric, i.e., if the identity
\begin{align*}\label{lsa1}
(x,y,z) & = (y,x,z)
\end{align*}
is satisfied for all $x,y,z \in A$. The algebra is called {\it RSA}, if
the identity
\begin{align*}
(x,y,z) & = (x,z,y)
\end{align*}
is satisfied.
\end{defi}

The opposite algebra of an LSA is an RSA. Indeed, if $x\cdot y$ is the product
in $A$, then $x\circ y=y\cdot x$ is the product in $A^{op}$.
An associative product is right- and left-symmetric. Note that the converse
is not true in general: the algebra $A:=Kx\oplus Ky$ with product
$x.x=0,\; x.y=0,\; y.x=-x,\; y.y=x-y$ is an RSA and LSA, but we have
$(y.y).y-y.(y.y)=x$. We note that LSAs and RSAs are examples of Lie-admissible algebras, i.e.,
the commutator
\begin{align*}
[x,y] & = x\cdot y-y\cdot x
\end{align*}
defines a Lie bracket. This follows from the identity 
\vspace*{0.5cm}
\begin{center}
\begin{tabular}{c}
$[[a,b],c]+[[b,c],a]+[[c,a],b] = $ \\[0.3cm]
$(a,b,c)+(b,c,a)+(c,a,b)-(b,a,c)-(a,c,b)-(c,b,a).$  
\end{tabular}
\end{center}
\vspace*{0.5cm}
valid in any $K$-algebra. We denote the Lie algebra by $\Lg_A$.

\section{Origins of left-symmetric algebras}

\subsection{Vector fields and RSAs}

Let $U$ be an associative commutative algebra, and 
$\CD=\{\partial_1,\ldots ,\partial_n\}$ be a system of commuting derivations of $U$.
If we regard the derivations in the endomorphism algebra we will require them to
be linearly independent. For any $u\in U$ the endomorphisms 
\[
u\partial_i \colon U \ra U, \quad (u\partial_i)(v)=u \partial_i(v)
\]
are derivations of $U$. Denote by $U\CD={\rm Vec}(n)$ the vector space
of derivations
\[
{\rm Vec}(n)=\left \{ \sum_{i=1}^n u_i\partial_i \mid u_i \in U, \partial_i \in \CD \right \}.
\]
We may consider this as a space of vector fields. We introduce on ${\rm Vec}(n)$ the following
algebra product
\begin{align}\label{vec}
u\partial_i \circ v \partial j & = v\partial_j(u)\, \partial_i .
\end{align}
 
\begin{prop}
The algebra $({\rm Vec}(n),\circ)$ is an RSA. It is called right-symmetric Witt algebra
generated by $U$ and $\CD$.
\end{prop}

\begin{proof}
The associator is given by
\begin{align*}
(u\partial_i,v\partial_j,w\partial_k) & = (u\partial_i \circ v \partial_j) \circ w \partial_k
- u\partial_i \circ (v \partial_j \circ w \partial_k) \\
 & = v \partial_j(u)\partial_i \circ w\partial_k-u\partial_i \circ (w\partial_k(v)\partial_j) \\
 & = w\partial_k(v\partial_j(u))\, \partial_i - w\partial_k(v)\partial_j(u)\, \partial_i \\
 & = w\{\partial_k(v)\partial_j(u)+v \partial_k(\partial_j(u)))\}\, \partial_i -  w\partial_k(v)
\partial_j(u)\, \partial_i \\
 & = wv\partial_k(\partial_j(u))\, \partial_i .
\end{align*}
Since $wv=vw$ in $U$ and the elements in $\CD$ commute it follows
\begin{align*}
(u\partial_i,w\partial_k,v\partial_j) & = vw  \partial_j(\partial_k(u))\, \partial_i \\
 & = (u\partial_i,v\partial_j,w\partial_k). 
\end{align*}
\end{proof}

As an example, let $M^n$ be a smooth $n$-dimensional flat manifold, $U$ be the algebra
of smooth functions on $M$ and ${\rm Diff}(n)$ the algebra of $n$-dimensional
differential operators 
\begin{align*}
\sum_{\al \in \Z^n}\la_{\al}u_{\al}\partial^{\al}, \quad \partial^{\al} =\prod_{i=1}^n \partial_i^{\al_i}
\end{align*}
where $\partial_i=\partial/\partial x_i$, $\al=(\al_1,\ldots ,\al_n)\in \Z^n$ and $u_{\al}\in U$.
Note that $\partial_i \partial_j=\partial_j\partial_i$ for all $i,j$ since $M^n$ is flat.
The subspace of differential operators of first order is just ${\rm Vec}(n)$ as above. 
It can be interpretated as as a space of vector fields on $M$. 
The algebra ${\rm Diff}(n)$ is associative, whereas the algebra ${\rm Vec}(n)$ under the product
\eqref{vec} is right-symmetric but not associative. \\
Other important special cases of $U$ are the polynomial ring $U=K[x_1,\ldots ,x_n]$
in $n$ variables, or the Laurent polynomial algebra $U=K[x_1^{\pm 1},\ldots ,x_n^{\pm 1}]$.
In the first case $U\CD$ is denoted by $W^r_n$ with Lie algebra $W_n$, the Witt algebra of rank $n$.
There exists a grading and a filtration of $U\CD$ as an RSA and as a Lie algebra.
For $n=1$ the algebra $W_1^r$ satisfies an additional identity:
\begin{align}\label{novikov}
x\circ (y\circ z) & = y \circ (x\circ z).
\end{align}
This means that the left multiplications in this algebra commute.
RSAs satisfying this identity are called {\it right Novikov algebras}.
There exists a large literature on Novikov algebras, see \cite{BAN}, \cite{BAM},  \cite{BU3}, \cite{OS1}, 
\cite{OS2}, \cite{ZEL} and the references given therein. 
For details concerning right-symmetric Witt algebras see \cite{DZL}and \cite{DZ1}.

\subsection{Rooted tree algebras and RSAs}

Probably Caley was the first one to consider RSAs. In his paper \cite{CAY} he also 
described a realization of the right-symmetric Witt algebra as a rooted tree algebra. \\
A {\it rooted tree} is a pair $(T,v)$ where $T$ is a non-empty finite, connected  
graph without loops and $v$ is a distinguished vertex of $T$ called the root.
This root gives an orientation of the graph; edges are oriented towards the root. 
Denote by $\abs{T}$ the set of vertices of $T$. 

Now we will introduce an algebra product on the vector space of rooted trees: 
Denote by $T_1\circ_v T_2$ the graph defined by adding to the disjoint union of $T_1$ and $T_2$ an edge joining
the root of $T_2$ to the vertex $v$ of $T_1$, and keeping the root of $T_1$ as the root.
In other words, for rooted trees $(T_1,v_1)$ and $(T_2,v_2)$ we have the rooted tree
$(T_1\circ_v T_2,v_1)$. Then we define
\[
(T_1,v_1) \circ (T_2,v_2) = \sum_{v\in \abs{T_1}} (T_1\circ_v T_2,v_1).
\]
The graph $T_1 \circ T_2$ is obtained by the sum over all possible graftings:
add a new branch to the root of $T_2$ and plant this graph to each node of $T_1$ and 
add the resulting trees. Here is an example:

\begin{center}
\scalebox{0.7}{\includegraphics{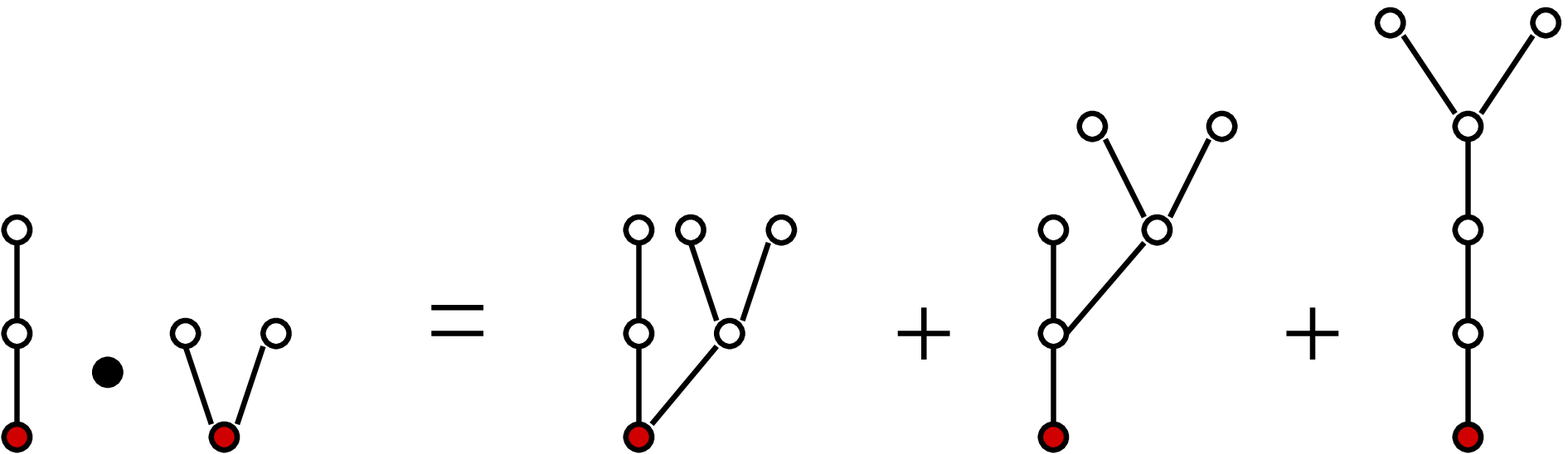}}
\end{center}

We have the following result \cite{CHL}:

\begin{prop}
The above product is  right-symmetric. The free RSA on a generator $\{ u\}$ has
the rooted trees as a basis.
\end{prop}

The right-symmetry for the associator of three elements
$(T_1,v_1),(T_2,v_2)$ , $(T_3,v_3)$ may be seen from the fact that insertion of
graphs is a local operation, and that on both sides, the difference amounts
to plugging the subgraphs $T_2,T_3$ into $T_1$ at disjoint places, which is
evidently symmetric under the exchange of $T_2$ and $T_3$. \\
In  \cite{DZL} the algebra of labelled rooted trees is considered.
Let $S$ be a set. A {\it labeling} of $T$ by $S$ is a map $\abs{T} \ra S$.
Denote the set of rooted trees labelled by $S$ by $T(S)$.
Identify two rooted labeled trees if there is an isomorphism of labelled graphs sending the root
to the root. For example, the following two labelled trees belong to the same class:

\begin{center}
\scalebox{0.7}{\includegraphics{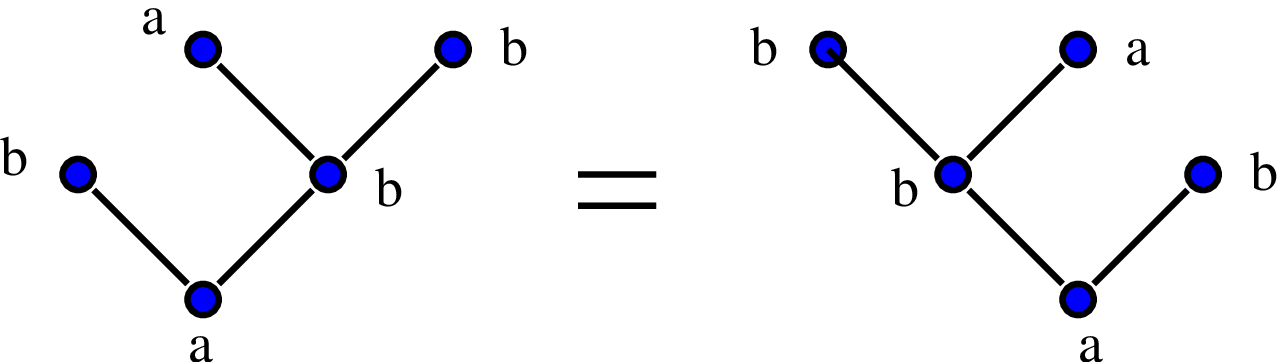}}
\end{center}

This is written as $T(a,b,T(b,a,b))=T(a,T(b,b,a),b)$.
Let us write a rooted tree as $T(v,x_1,\ldots ,x_n)$, where $x_i$ are trees,
$v$ is the root and $n$ is the number of incomming edges of the root.
Define a non-associative and non-commutative operation on $T(S)$ by
\[
T(v,x_1,\ldots ,x_n)\bullet y = T(v,x_1,\ldots ,x_n,y).
\]
This satisfies the identity $(a\bullet b)\bullet c=(a\bullet c)\bullet b$.\\
Let $R$ be a commutative ring. Define the {\it tree algebra} $\CT(S)$ as the
free $R$-module on $T(S)$. A bilinear multiplication $\circ$ on $\CT(S)$ is
defined recursively on basis elements as follows. Let $v\in S$ and $x_1,\ldots,x_n,y\in T(S)$. Then
\begin{align*}
v\circ y & = v\bullet y =T(v,y), \\
T(v,x_1,\ldots ,x_n)\circ y & = T(v,x_1,\ldots ,x_n,y) \\
&  + \sum_{i=1}^{n} T(v,x_1,\ldots ,\widehat{x_i}, \ldots , x_n)\bullet (x_i\circ y).
\end{align*}
It follows that $(\CT(S),\circ)$ is a derivation algebra of $(\CT(S),\bullet)$:
\[
(x\bullet y)\circ z = (x\circ z)\bullet y+x \bullet (y\circ z).
\]
Moreover, $(\CT(S),\circ)$ is right-symmetric. We have the following result \cite{DZL}:

\begin{prop}
As an $R$-algebra, $\CT(S)$ is generated by $S$.
\end{prop}

Let $\Lt (S)$ be the Lie algebra of $\CT(S)$ and $\CH_S=(U(\Lt(S))^{\ast}$ be the the dual
of the universal enveloping algebra of $\Lt (S)$. The algebra $\CH_S$ is a Hopf algebra.

\begin{rem}
Hopf algebras and RSAs of rooted trees play an important role in renormalizable
quantum field theories. Feynman graphs form an RSA.
\end{rem}

More precisely, for any QFT the combinatorics of Feynman graphs gives rise to an
RSA of rooted trees, and hence to a Lie algebra and a Hopf algebra of rooted trees.
In fact, the structure of the pertubative expansion of a QFT is in many ways
determined by the Hopf and Lie algebra structures of Feynman graphs.
This allows a conceptual interpretation of renormalization theory. For example,
the Hopf algebra of rooted trees yields the finite renormalization needed to
satisfy the requirements of quantized gauge symmetries. There is an extensive
literature available, see \cite{COK}, \cite{KR1}, \cite{KR2} and the references given
therein.\\
One should note that it is possible to construct on any class $\CC$ of graphs
a right-symmetric product: for each pair of graphs $\Ga_1,\Ga_2 \in \CC$ one
defines a set $I(\Ga_1,\Ga_2)$ and a map 
\[
\ga \colon I(\Ga_1,\Ga_2) \ra \CC
\]
where $I(\Ga_1,\Ga_2)$ is, roughly spoken, the set of possible insertions of
$\Ga_2$ in $\Ga_1$ staying in the class $\CC$, and $\ga$ realizes these insertions.
Then the product
\[
\Ga_1 \star \Ga_2 = \sum_{i\in I(\Ga_1,\Ga_2)} \ga (i)
\]
is right-symmetric, i.e., $\Ga_1\star (\Ga_2\star \Ga_3)- (\Ga_1\star \Ga_2)\star \Ga_3$
is symmetric in $\Ga_2$ and $\Ga_3$. To see this, one has to prove 
that this associator corresponds to the insertion of  $\Ga_2$ and $\Ga_3$ at two distinct 
vertices of $\Ga_1$, which is of course symmetric in  $\Ga_2$ and $\Ga_3$.
This way it is possible to consider also classes of graphs with certain
constraints, e.g., with renormalization conditions.

\subsection{Words in two letters and RSAs}

The right-symmetric structure on certain graphs can also be illustrated by words on an
alphabet. We want to consider the following nice construction.
Let $W$ be the vector space generated by the set of finite words on the alphabet
$\{A,B\}$. Let $\emptyset$ denote the empty word. If $x$ is such a word, then
let $x[i]$ denote the $i$-th letter of $x$. For example, if $x=AB^2AB$ then
$x[0]=\emptyset$ and $x[4]=A$. Let $\ell(x)$ be the length of the word $x$.
Define an algebra product on $W$ by the formula
\[
x\circ y = \sum_{i=0}^{\ell(x)}\ep (i) x \curvearrowleft ^i y
\]
where $x \curvearrowleft ^i y$ is the insertion of $y$ between $x[i]$ and $x[i+1]$
and 

$$
\ep (i) =
\begin{cases}
-1 & \text{ if } x[i]=A \text{ and } x[i+1]=B \\
+1 & \text{ if } x[i]=B \text{ and } x[i+1]=A \text{ or } \emptyset\\
+1 & \text{ if } x[i]=\emptyset \text{ and } x[i+1]=A \\
0  & \; \text{else}
\end{cases}
$$
Note that nothing is inserted between $A$ and $\emptyset$, or
between $\emptyset$ and $B$.

\begin{ex}
Let us compute a few examples:
\begin{align*}
A \circ A & = A^2 \\
A \circ AB & = ABA  \\
AB \circ A & = A^2B-A^2B+ABA = ABA \\
ABA \circ B & = BABA \\
BA \circ AB & = BABA \\
AB \circ AB & = 2ABAB-A^2B^2 \\
BA \circ ABA & = BABA^2 \\
ABA \circ BA & = BA^2BA-ABABA+AB^2A^2
\end{align*}
\end{ex}

The product is neither commutative nor associative. Indeed,
\begin{align*}
(ABA \circ B)\circ BA & = BABA \circ BA \\
 & = B^2A^2BA-BABABA+BAB^2A^2,
\end{align*}
whereas
\begin{align*}
ABA \circ (B\circ BA) & = ABA\circ B^2A \\
 & = B^2A^2BA-AB^2ABA+AB^3A^2.
\end{align*}
It follows that
\begin{align*}
ABA \circ (B\circ BA)-(ABA \circ B)\circ BA & = \\
-BABABA+BAB^2A^2+AB^2ABA-AB^2A^2 & = \\
(ABA \circ BA)\circ B - ABA \circ (BA\circ B).
\end{align*}
Hence the associators satisfy
\[
(ABA,B,BA)=(ABA,BA,B).
\]
This is no coincidence. We have the following result:

\begin{prop}
The algebra $(W,\circ)$ is right-symmetric, i.e.,
\[  
x\circ (y\circ z) - (x\circ y)\circ z = x\circ (z\circ y) - (x\circ z)\circ y  
\]
for all words $x,y,z$ in $W$.
\end{prop}

\subsection{Vertex algebras and LSAs}

Vertex algebras have been studied very intensively over the last years. There is a huge
literature on this subject. We can only mention just a few classical references here:
\cite{FHL}, \cite{FLM}, \cite{KAC}, \cite{LEL}.
We will try to explain what a vertex algebra is, and how it is related to LSAs.
Vertex algebras were first introduced by R. Borcherds in $1986$, see \cite{BOR}.
The definition is given in terms of quite complicated identities, the so called
Borcherds identities. 
In $1996$ Kac \cite{KAC} gave an equivalent definition of a vertex
algebra as a pointed vector space $(V, |0\rangle)$ together with a local state-field
correspondence $Y$. Here any vector space $V$ with a fixed non-zero vector $|0\rangle$,
referred as the vacuum vector, is called a pointed vector space. 
Kac also introduced conformal algebras, which can be defined easily. Then he proved, 
using field algebras, that vertex algebras form a subclass of conformal algebras, see \cite{BBK}.
This allows to give an easier definition of vertex algebras, with fewer axioms.
This goes as follows:

\begin{defi}
A {\it Lie conformal algebra} is a $\C[T]$-module $V$
endowed with a $\C$-linear map $V\otimes V\ra \C[\la]\otimes V$ denoted by
$a\otimes b\mapsto [a_{\la}b]$, called the {\it $\la$-bracket}, satisfying the following
axioms for all $a,b,c \in V$:
\begin{align*}
[(Ta)_{\la}b]  & = -\la [a_{\la}b],\\
[a_{\la}(T b)] & = (\la+T) [a_{\la}b],\\
[b_{\la}a] & = [a_{-\la-T}b],\\
[[a_{\la}b]_{\la+\mu} c] & = [a_{\la}[b_{\mu}c]-[b_{\mu}[a_{\la}c]].
\end{align*}
\end{defi} 
The first two axioms are called {\it sesquilinearity}.
Together they say that $T$ is a derivation of the $\la$-bracket: $T[a_{\la}b]=[(Ta)_{\la}b]+[a_{\la}(T b)]$.
The third axiom is called {\it skewsymmetry}, and the last one the {\it Jacobi identity}.
Now the theorem in \cite{BBK} is as follows:

\begin{thm}
Giving a vertex algebra structure on a pointed vector space $(V,|0\rangle)$ is the
same as providing $V$ with the structures of a Lie $\C[T]$-conformal algebra and a left-symmetric
$\C[T]$-differential algebra with unit $|0\rangle$, satisfying  
\begin{align*}
a.b - b.a & = \int_{-T}^0 d\la\; [a_{\la} b].\\
[a_{\la} (b.c)] & = [a_{\la}b]. c+ b. [a_{\la}c]+\int_0^{\la}d\mu \;
[[a_{\la}b]_{\mu}c].
\end{align*}
\end{thm}

The first axiom is called {\it skewsymmetry} and the second
is called the {\it non-commutative Wick formula}. 
The theorem says that we can define a vertex algebra as follows:

\begin{defi}
A {\it Vertex algebra} is a pair $(V, |0\rangle)$, where $V$ is a $\C[T]$-module
and $|0\rangle$ is an element of $V$ (the vacuum state), endowed with two operations: a
$\la$-bracket $V\otimes V\ra \C[\la]\otimes V$, $a\otimes b\mapsto [a_{\la}b]$
making it a Lie conformal algebra, and a {\it normally ordered product} $V\otimes V\ra V$,
$a\otimes b\mapsto a.b$, which makes it a unital differential algebra with unit  $|0\rangle$
and derivation $T$. These two operations satisfy 
the following axioms:\\
\begin{align*}
(a. b). c -a. (b. c) & = \left( \int_0^T d\la\; a \right). [b_{\la}c]+
\left( \int_0^T d\la\; b \right).  [a_{\la}c],\\
a. b- b. a & = \int_{-T}^0 d\la\; [a_{\la} b],\\
[a_{\la} (b.c)] & = [a_{\la}b]. c+ b. [a_{\la}c]+\int_0^{\la}d\mu \;
[[a_{\la}b]_{\mu}c].
\end{align*}
\end{defi}

The first axiom here is called {\it quasi-associativity}. 
It follows that the underlying algebra of a vertex algebra is an LSA: indeed,
the right-hand side of the first identity is symmetric with respect to $a$ and $b$.
Hence the product $a.b$ is left-symmetric. More details can be found also in \cite{ROS}.
For any Lie conformal algebra $R$ one can construct a so called enveloping vertex
algebra $U(R)$. Hence each example of a Lie conformal algebra produces an example of a vertex algebra.

\begin{ex}
The Virasoro Lie conformal algebra is given by
\begin{align*}
V & = \C[T]L\oplus \C\, |0\rangle 
\end{align*}
with $\la$-bracket 
\begin{align*}
[L_{\la}L] & = (T+2\la)L+\frac{c}{12}\la^3 |0\rangle ,
\end{align*}
where $c\in \C$ is the central charge.
\end{ex}

\begin{rem}
Finite-dimensional simple Lie conformal algebras have been classified, see \cite{DEK}
and the references cited therein. For infinite-dimensional algebras this classification 
is far from being solved.
\end{rem}

\subsection{Operad theory and RSAs}

Let $\CS_n$ denote the symmetric group on $n$ letters and $K[\CS_n]$ the
group ring. For us an {\it operad} $\CP$ is a sequence of $K[\CS_n]$-modules
$(\CP(n))_{n\ge 1}$ equipped with a unit $1 \in P(1)$, together with composition products, for
$n,m_1,\ldots ,m_m \in \N$,
\[
\ga \colon \CP(n)\otimes \CP(m_1)\otimes \cdots \otimes  \CP(m_n) \ra P(n+m_1+\cdots +m_n), 
\]
satisfying natural associativity, unitarity and equivariance conditions. For details see \cite{MAY}.
There is a natural grading on the total space $\oplus_n \CP(n)$, defined by
\[
\Bigl( \bigoplus_{n\ge 1} \CP(n)\Bigr)^m=\CP(m+1).
\]

\begin{ex}
Let $V$ be a $K$-vector space and define 
\[
\CP(n)=\Hom_K (V^{\otimes n},V).
\]
Then $\CP=\CE nd (V)=(\CP(n))_{n\ge 1}$ forms an operad. 
\end{ex}
The $\CS_n$-actions are given by permutations of tensors on $V^{\otimes n}$. 
The compositions are the usual ones for multilinear maps. \\
For $p\in \CP(n)$ and $q\in \CP(m)$ denote
\[
p \circ_i q = \ga (p,\id^{\otimes i-1},q,\id^{\otimes n-i}).
\]
Then we have the following result:
\begin{prop}
Let $\CP$ an operad of vector spaces. Then the graded vector space
$\bigoplus_{n\ge 1} \CP(n)$ forms an RSA under the product
\[
p\circ q = \sum_{i=1}^n  p \circ_i q 
\]
where  $p\in \CP(n)$ and $q\in \CP(m)$.
\end{prop}
Recall the notation $\CT(n)$ for the free $\Z$-module of rooted trees labelled
by $S=\{ 1,\ldots ,n \}$. We can endow $\CP\CR=(\CT(n))_{n\ge 1}$ with an operad
structure by defining compositions $T\circ_i S$ using substitutions and graftings
in a certain way. For more details see \cite{CHL}.
On the other hand we have the quadratic binary operad $\CP\CL$ defining RSAs.
One constructs this operad as follows. Let $\CF$ be the free operad generated by
the regular representation of $\CS_2$. A basis of $\CF(n)$, as a vector space,
is given by products $(x_{i_1}x_{i_2}\ldots x_{i_n})$ indexed by $\{ 1,\ldots ,n\}$
with arbitrary bracketing. For instance, a basis of $\CF(2)$ is given by the products
$(x_1x_2)$ and  $(x_2x_1)$; and a basis of $\CF(3)$ is given by the products
$((x_1x_2)x_3),\, (x_1(x_2x_3))$ and all their permutations. Then
$\CP\CL=\CF/I$ where $I$ denotes the ideal of $\CF$ generated by the 
$\CS_3$-submodule of $\CF(3)$ given by the relation
\[
((x_1x_2)x_3)-(x_1(x_2x_3))-((x_1x_3)x_2)+(x_1(x_3x_2)).
\] 
The following result was proved in  \cite{CHL}:
\begin{prop}
The operad  $\CP\CL$ defining RSAs is isomorphic to the operad  $\CP\CR$ of
rooted trees.
\end{prop}

\subsection{Deformation complexes of algebras and RSAs}

Let $V$ be an $R$-module and denote by $C^n(V,V)$ the space of all $n$-multilinear
maps from $V$ to $V$. For $f\in C^p(V,V)$ and $g\in C^q(V,V)$ consider
the product 
\[
\kringel \colon C^p(V,V)\times C^q(V,V) \ra C^{p+q-1}(V,V), \;
(f,g) \mapsto f\kringel g
\] 
given by
\begin{align*}
(f\kringel g)(x_1,\ldots ,x_{p+q-1}) & = \sum_{i=1}^p f(x_1,\ldots , x_{i-1},
g(x_i,\ldots , x_{i+q-1}), x_{i+q},\ldots , x_{p+q-1}).
\end{align*}
Let us denote the $i$-th summand by $f\kringel_i g$. One can show
that this product is indeed right-symmetric:

\begin{prop}
The algebra $(C^{\bullet}(V,V),\kringel)$ is an RSA.
\end{prop}

Gerstenhaber \cite{GER} already noted this fact in a graded version which arises in the
Hochschild cohomology setting. Let $A$ be an associative algebra and 
$C^n(A,A)=\Hom_K(A^{\otimes n},A)$ be the space of Hochschild $n$-cochains. Then the main tool 
in studying the deformation theory of $A$ is the Hochschild complex
\[
0 \ra C^0(A,A)\xrightarrow{d} \cdots \xrightarrow{d} C^n(A,A) \xrightarrow{d} C^{n+1}(A,A) 
\xrightarrow{d} \cdots
\]
denoted by $C^{\bullet}(A,A)$. 
Gerstenhaber defined a product on this complex as follows: 
\begin{align*}
(f\circ g)(x_1,\ldots ,x_{p+q-1}) & = \sum_{i=1}^p (-1)^{(q-1)(i-1)} (f\kringel_i g)(x_1,\ldots ,
x_{p+q-1}).
\end{align*}
This is a graded version of the product given above. It is also not associative in general,
but satisfies a graded right-symmetric identity.

\begin{defi}
Let $V$ be a graded vector space and $\abs{x}$ denotes the degree of
$x\in V$. Then $V$ together with a $K$-bilinear product $(x,y)\ra x\cdot y$
is called a {\it graded RSA}, if
\begin{align*}
(x\cdot y)\cdot z - x\cdot (y\cdot z) & = (-1)^{\abs{y}\abs{z}}((x\cdot z)\cdot y - x\cdot (z\cdot y)).
\end{align*}
\end{defi}

We have the following result, see  \cite{GER}, \cite{NI1}:

\begin{prop}
The algebra $(C^{\bullet}(A,A),\circ)$ is a graded RSA.
\end{prop}

The composition bracket
\begin{align*}
\llbracket x,y \rrbracket & = x\circ y-(-1)^{\abs{x}\abs{y}} y\circ x
\end{align*}
is a graded Lie bracket, called the {\it Gerstenhaber bracket}. It is graded
skew-symmetric, i.e., 
\begin{align*}
\llbracket x,y \rrbracket & = -(-1)^{\abs{x}\abs{y}} \llbracket y,x\rrbracket,
\end{align*}
and satisfies the graded Jacobi identity
\begin{align*}
(-1)^{\abs{x}\abs{z}}\llbracket \llbracket x,y\rrbracket,z\rrbracket
+(-1)^{\abs{y}\abs{x}}\llbracket \llbracket y,z\rrbracket,x\rrbracket
+(-1)^{\abs{z}\abs{y}} \llbracket \llbracket z,x\rrbracket,y\rrbracket & = 0.
\end{align*}
Note that the Hochschild coboundary map $d$ satisfies $d(f)=-\llbracket \mu,f\rrbracket $, where
$\mu \in \Hom_K(A\otimes A,A)$ is the multiplication map of $A$.

\subsection{Convex homogeneous cones and LSAs}

Convex homogeneous cones arose in the theory of automorphic functions on
bounded homogeneous domains in $\C^n$. If $V$ is a convex homogeneous cone
in $\R^n$ then the domain $D=\{x+iy \mid y \in V\}\subseteq \C^n$ is
analytically equivalent to a bounded homogeneous domain. This is the so-called
{\it generalized upper half-plane}, or {\it Siegel domain of the first kind}.
It is homogeneous with respect to the group of complex affine transformations
of $D$.

\begin{defi}
A {\it convex cone} in $\R^n$ is a non-empty set $V$ having the following properties:
\begin{itemize}
\item[(1)] if $x\in V$ and $\la >0$ then $\la x\in V$;
\item[(2)] if $x,y \in V$ then $x+y \in V$;
\item[(3)] the closure of $V$ does not contain a subspace of positive dimension;
\item[(4)] the set $V$ is open in $\R^n$.
\end{itemize}
\end{defi}

Condition $(3)$ says that $V$ does not completely contain any straight line.
The subgroup $G(V)$ of $GL(\R^n)$ consisting of the automorphisms $A$ 
satisfying $AV=V$ is called the {\it automorphism group of $V$}. A convex
cone $V$ is called {\it homogeneous} if $G(V)$ acts transitively on it. 
As an example consider the cone of positive-definite symmetric matrices
in $M_n(\R)$, or the cone of positive-definite Hermitian matrices in
$M_n(\C)$.

\begin{defi}
A {\it convex domain} in an affine space $P$ is any nonempty open convex set
$U\subset P$ not completely containing any straight line.
\end{defi}

Clearly, a convex cone is a special case of a convex domain. 
The vertex of the cone defines an origin in the affine space and converts it
into a linear space.
The group of affine transformations leaving $U$ invariant is denoted by $G(U)$.
It is an algebraic group. Let $\Lg(U)$ be its Lie algebra. 
We have $G(U)=K(U)T(U)$ and $K(U) \cap T(U)= \{e\}$, where $K(U)$ is the 
stability subgroup of some point $x_0\in U$ and $T(U)$ is a maximal connected 
triangular subgroup of $G(U)$. The group $T(U)$ acts simply transitively on 
$U$ by affine transformations. Let $\Lt (U)$ denote its Lie algebra. Let
$D\in \Lt(U)$, $x_0\in U$. Then the mapping
\[
D \ra D(x_0)
\]
is an isomorphism of the linear space $T(U)$ onto the linear space $R_P$ of free 
vectors of $P$. Let $D_a$ be the inverse image of the vector $a\in R_p$ under this mapping,
i.e., $D_a(x_0)=a$. Let $L_a$ denote the linear part of $D_a$ and define a bilinear
product on $R_P$ by
\begin{align*}
a\cdot b & = L_a(b).
\end{align*}
This algebra $(R_P,\cdot)$ is called the {\it algebra of $U$} with respect to the point $x_0$
and the group $T(U)$. Different choices of $x_0$ and $T(U)$ would lead to isomorphic
algebras, so we may speak of {\it the} algebra of $U$.
We have the following result, see \cite{VIN}:

\begin{thm}
The algebra $(R_P,\cdot)$ of any convex homogeneous domain is a left-symmetric algebra
over $\R$ satisfying the following properties:
\begin{itemize}
\item[(1)] there exists a linear form $s$ on $R_P$ such that $s(a\cdot b)=s(b\cdot a)$ and
$s(a\cdot a)>0$ for each $a\neq 0$.
\item[(2)] the eigenvalues of the operators $L_a\colon x\ra a\cdot x$ are real.
\end{itemize}
\end{thm}
It follows from the commutation rule of elements in $\Lg(U)$ that
\begin{align*}
[D_a,D_b](x_0) & = L_a(b)-L_b(a)=a\cdot b - b\cdot a,\\
[D_a,D_b] & = D_{a\cdot b-b\cdot a}, \\
[L_a,L_b] & =L_{a\cdot b-b\cdot a}
\end{align*}
This implies that we have $(a,b,c)=(b,a,c)$. The linear form is given by $s(a)=\tr (L_a)$.
Since $0=\tr([L_a,L_b])=\tr(L_{a\cdot b-b\cdot a})$ we have $s(a\cdot b)=s(b\cdot a)$.
Since the group $T(U)$ is triangular, the linear translations $L_a$ are simultaneously
reducible to triangular form and have real eigenvalues. In the special case that
$U$ is a convex homogeneous cone we obtain the following result:

\begin{cor}
If $U$ is a convex homogeneous cone then the algebra  $R_P$ has in addition
a two-sided unit element, i.e.,
\begin{itemize}
\item[(3)] there exists an element $e$ such that $e\cdot a=a\cdot e=a$ for all $a\in R_P$.
\end{itemize}
\end{cor}

Vinberg \cite{VIN} called LSAs satisfying the conditions $(1)$ and $(2)$ {\it clans}.
He described how to construct a convex homogeneous domain from a clan. 
This leads to the following result:

\begin{thm}
There is a one-to-one correspondence of $n$-dimensional convex homogeneous cones and
$n$-dimensional LSAs satisfying $(1),(2),(3)$.
\end{thm}

There exists a certain classification of this special class of LSAs, i.e., of clans with
unity. According to Vinberg this classification does not have the definite nature 
of, say, the classification of semisimple Lie algebras. More details are to be found in
\cite{VIN}, \cite{DOR} and the references given therein.

\subsection{Affine manifolds and LSAs}

Let $G$ be a Lie group acting smoothly and transitively on a
smooth manifold $X$.
Let $U\subset X$ be an open set and let $f: U\rightarrow X$ be a
smooth map. The map $f$ is called {\it locally}--$(X,G)$ if for each component
$U_i\subset U$, there exists $g_i\in G$ such that the restriction
of $g_i$ to $U_i\subset X$ equals the restriction of $f$ to $U_i\subset U$.

\begin{defi}
Let $M$ be a smooth manifold of the same dimension as $X$.
An {\it $(X,G)$--atlas} on $M$ is a pair $(U,\Phi)$ where $U$ is an open
covering of $M$ and $\Phi=\{\phi_{\al}: U_{\al} \rightarrow X \}_{U_{\al}\in
U}$ is a collection of coordinate charts such that for each pair
$(U_{\al},U_{\be})\in U\times U$ the restriction of
$\phi_{\al}\kringel \phi_{\be}^{-1}$ to $\phi_{\be}(U_{\al}\cap U_{\be})$ is
locally--$(X,G)$. An {\it $(X,G)$--structure on
$M$} is a maximal $(X,G)$--atlas and $M$ together with an $(X,G)$--structure
is called an {\it $(X,G)$--manifold}.
\end{defi}

Let $\Aff (\R^n)$ be the group of affine transformations which is given by
$$\left \{ \begin{pmatrix} A & b \\ 0 & 1 \end{pmatrix} \mid A \in GL_n(\R),\,
b\in \R^n \right \}.$$
It acts on the real affine space $\{(v,1)^t \mid v\in \R^n\}$ by
$$ \begin{pmatrix} A & b \\ 0 & 1 \end{pmatrix} \begin{pmatrix} v \\ 1
\end{pmatrix}=\begin{pmatrix} Av+ b \\ 1 \end{pmatrix}$$

\begin{defi}
Let $M$ be an $n$-dimensional manifold. An $(X,G)$--struc\-ture on $M$, where
$X$ is the real $n$--dimensional affine space, also denoted by
$\R^n$ here, and $G=\Aff (\R^n)$ is called an
{\it affine structure} on $M$ and $M$ is called an {\it affine manifold}.
\end{defi}

Affine structures on a smooth manifold $M$ are in correspondence with a
certain class of connections on the tangent bundle of $M$.
The following result can be found in \cite{KOB}:

\begin{prop}
There is a bijective correspondence of affine structures on a manifold $M$ and
flat torsionfree affine connections $\nabla$ on $M$.
\end{prop}

Let $\LX$ denote the Lie algebra of all differentiable vector fields on $M$.
The affine connection $\nabla$ is called {\it torsionfree}, or {\it symmetric} if
\begin{equation}\label{tors}
\nabla_X(Y)-\nabla_Y(X)-[X,Y]=0,
\end{equation}
and {\it flat} or of {\it curvature zero}, if
\begin{equation}\label{flat}
\nabla_X\nabla_Y-\nabla_Y\nabla_X-\nabla_{[X,Y]}=0.
\end{equation}
Such a connection determines a covariant differentiation 
\[
\nabla_X \colon \LX \ra \LX, \; \nabla_X : Y \mapsto \nabla_X (Y)
\]
for vector fields $X,Y \in \LX$. If we put
\[
X\cdot Y =  \nabla_X (Y)
\]
then we obtain an $\R$-bilinear product on $\LX$. The vanishing of
curvature and torsion, i.e. \eqref{tors} and \eqref{flat} is
equivalent to the following identities:

\begin{align}
[X,Y] & = X\cdot Y - Y \cdot X \\
[X,Y]\cdot Z & = X\cdot (Y\cdot Z)-Y\cdot (X\cdot Z)
\end{align}
\smallskip\noindent
Thus the given product makes $\LX$ into an LSA. \\
When do affine structures exist on a manifold $M$ ?
A flat Euclidean structure on a manifold automatically gives an affine
structure. It is well known that the torus and the Klein bottle
are the only compact two-dimensional manifolds that can be given
Euclidean structures \cite{THU}. 
Let $M$ be a closed $2$--manifold, i.e., compact and without boundary. 
If $M$ is a $2$--torus, then there exist many
affine structures, among them {\it non-Euclidean} ones.
A classification of all affine structures on the $2$--torus is given in
\cite{KUI},\cite{NAY}. If $M$ is a closed $2$--manifold 
different from a $2$--torus or the Klein bottle, then
there exist no affine structures. This
follows from Benzecri's result \cite{BEZ} of $1955$:
\begin{thm}
A closed surface admits affine structures if and only if its Euler
characteristic vanishes.
\end{thm}
In higher dimensions there is no such criterion for the
existence of an affine structure. However, Smillie \cite{SMI} proved
that a closed manifold does not admit an affine structure
if its fundamental group is built up out of finite groups by taking
free products, direct products and finite extensions.
In particular, a connected sum of closed manifolds with finite
fundamental groups admits no affine structure.
It is also known \cite{CDM} that certain Seifert fiber spaces admit no affine structure.
Let $M$ be a Seifert fiber space with vanishing first Betti number.
Then $M$ does not admit any affine structure.

\subsection{Left-invariant affine structures on Lie groups and LSAs}

Let $G$ be a connected and simply connected Lie group, with Lie algebra $\Lg$.

\begin{defi}
An affine structure on $G$ is called {\it left-invariant}, if each
left-multiplica\-tion map $L(g): G\rightarrow G$ is an affine diffeomorphism.
\end{defi}

If $\Gamma$ is a discrete subgroup of $G$ then the coset space $G/\Gamma$ 
inherits an an affine structure from $G$ by the left-invariance of the
structure. Many examples of affine manifolds can be constructed
via left-invariant affine structures on Lie groups. 

\begin{rem}
It is well known that $G$ admits a complete left-invariant affine structure if
and only if $G$ acts simply transitively by affine tranformations on $\R^n$.
Auslander proved that in this case $G$ is solvable \cite{AUS}.
\end{rem}

Milnor had posed in connection with Auslander's conjecture
on affine crystallographic groups, the following question in \cite{MIL}:

\begin{conm}
Does every solvable Lie group $G$ admit a complete left--inva\-ri\-ant affine
structure, or equivalently, does the universal covering group
$\widetilde{G}$ operate simply transitively by affine transformations
of $\R^k$ ?
\end{conm}

It is possible to formulate Milnor's problem in purely algebraic terms.

\begin{defi}\label{affine}
An {\it affine, or left-symmetric structure} on a Lie algebra
$\Lg$ is a $K$--bilinear product $\Lg \times \Lg \rightarrow \Lg$
which is left-symmetric and satisfies
\begin{equation}\label{lsa2}
[x,y]=x\cdot y -y\cdot x.
\end{equation}
\end{defi}
Denote the left-multiplication in the LSA
by $L(x)y=x\cdot y$, and the right multiplication by $R(x)y=y\cdot x$.

\begin{prop}\label{bi}
There are canonical one-to-one correspondences between the following classes
of objects, up to suitable equivalence:
\begin{itemize}
\item[(a)] \{Left-invariant affine structures on $G$\}
\item[(b)] \{Affine structures on the Lie algebra $\Lg$\}
\end{itemize}
Under the bijection, bi-invariant affine structures correspond to associative 
LSA--struc\-tures.
\end{prop}
\begin{proof}
The details of the correspondence are given in \cite{BU1} and \cite{DEM}.
Suppose $G$ admits a left-invariant affine structure. Then
there exists a left-invariant flat torsionfree affine connection $\nabla$
on $G$. Since $\nabla$ is left-invariant, for any two left-invariant vector
fields $X,Y\in \Lg$, the covariant derivative $\nabla_X(Y)\in \Lg$ is
also left-invariant. Hence covariant differentiation defines a bilinear
multiplication on $\Lg$ :
$$\Lg \times \Lg \ra \Lg,\; (X,Y)\mapsto XY=\nabla_X(Y).$$
The conditions that $\nabla$ has zero torsion and zero curvature
amounts as before to
\begin{align*}
XY-YX & =[X,Y], \\
X(YZ)-Y(XZ) & =[X,Y]Z=(XY)Z-(YX)Z.
\end{align*}
This multiplication is an affine structure on $\Lg$ by definition.
\end{proof}

Hence the algebraic version of Milnor's question is given as folllows:

\begin{conm}
Does every solvable Lie algebra admit a complete affine structure ?
\end{conm}

Milnor's question has a very remarkable history.
When he asked this question in $1977$, there was some evidence for the
existence of such structures. After that many articles appeared proving
some special cases, see for example \cite{AUS}, \cite{KIM}, \cite{SCH}.
However, the general question was
still open and it was rather a conjecture than a question by the time.
Many mathematicians believed that Milnor's question should have a positive
answer. In fact, around $1990$ there appeared articles in the
literature which claimed to prove the conjecture, e.g., \cite{BOY} and
\cite{NIS}. However, in $1993$ Yves Benoist constructed a counterexample in dimension
$11$ consisting of a filiform nilpotent Lie group without
any left-invariant affine structure. Almost at the same time
we have produced a whole family of counterexamples \cite{BU1}, \cite{BU4}, \cite{BU6} for the
dimensions $10\le n\le 13$, out of which Benoist's example ermerges
as just one in a series:

\begin{thm}
There are filiform nilpotent Lie groups of dimension 
$10\le n \le 13$ which do not admit any left-invariant affine structure.
Any filiform nilpotent Lie group of dimension $n\le 9$ admits
a left-invariant affine structure.
\end{thm}

For the proof see \cite{BU6}. An important role plays the following observation, see
Proposition $\ref{faithful}$: if $\Lg$ admits an affine structure then $\Lg$
possesses a faithful Lie algebra module of dimension $\dim \Lg+1$.

\begin{rem}
It seems that there exist counterexamples in all dimensions
$n\ge 10$. This is not proved yet. Moreover no good criteria are known to decide the
existence question for a given Lie group. We have suggested
in \cite{BU5} that the existence of affine structures on $\Lg$ in some cases
depends on the cohomology group $H^2(\Lg,K)$.
\end{rem}

\section{Algebraic theory of LSAs}

\subsection{Faithful representations and affine structures}

Let $A$ be a left-symmetric algebra over $K$ with underlying Lie algebra
$\Lg$. By definition the product $x\cdot y$ in $A$ satisfies the two
conditions
\begin{align*}
x\cdot (y\cdot z)-(x\cdot y)\cdot z & = y\cdot (x\cdot z)-(y\cdot x)\cdot z\\
[x,y] & = x\cdot y -y\cdot x
\end{align*}

for all $x,y,z \in A$. The left-multiplication $L$ in $A$ is given by
$L(x)(y)=x\cdot y$. The two conditions are equivalent to
\begin{gather}
L: \Lg \rightarrow \Lg\Ll (\Lg) \text{ is a Lie algebra homomorphism}\label{7}  \\ 
\I : \Lg \rightarrow \Lg_{L} \text{ is a $1$--cocycle in } Z^1(\Lg,\Lg_{L}) \label{8}
\end{gather}
where $\Lg_{L}$ denotes the $\Lg$--module with action given by $L$, and
$\I$ is the identity map. $Z^1(\Lg,\Lg_L)$ is the space of $1$--cocycles
with respect to $\Lg_L$. Note that the right-multiplication $R$ is in general not a
Lie algebra representation of $\Lg$.
Recall that, for a $\Lg$-module $M$, the space
of $1$-cocycles and the space of $1$-coboundaries is given by
\begin{align*}
Z^1(\Lg,M) & = \left \{ \om \in \Hom (\Lg,M) \mid \om ([x,y])=
x\pkt \om (y)-y\pkt \om(x) \right \}, \\
B^1(\Lg,M) & = \{ \om \in \Hom (\Lg,M) \mid \om (x)=x\pkt m
\mbox{ for some } m\in M \}.
\end{align*}

Let $\Lg$ be of dimension $n$ and identify $\Lg$ with $K^n$ by
choosing a $K$--basis. Then $\Lg\Ll(\Lg)$ gets identified with $\Lg \Ll_n(K)$.
\begin{defi}
The Lie algebra of the Lie group $\Aff (G)$ is called the Lie algebra of
{\it affine transformations} and is denoted by $\La \Lf\Lf (\Lg)$.
It can be identified as a vector space with $\Lg \Ll_n(K)\oplus K^n$.
\end{defi}
Given an affine structure on $\Lg$, define a map
$\al : \Lg \rightarrow \La\Lf\Lf(K^n)$ by $\al(x)=(L(x),x)$.
That is a Lie algebra homomorphism:
\begin{lem}\label{127}
The linear map $L \colon \Lg \rightarrow \Lg\Ll(\Lg)$ satisfies \eqref{7} and \eqref{8}
if and only if $\al : \Lg \rightarrow \La\Lf\Lf(K^n)$ is a Lie algebra homomorphism.
\end{lem}
\begin{proof}
Let more generally $\al(x)=(L(x),t(x)) \in
\Lg \Ll_n(K)\oplus K^n$ with a bijective linear map $t : \Lg
\rightarrow \Lg$. We have
\begin{equation}
\al([x,y])=[\al(x),\al(y)] \Longleftrightarrow \begin{cases}
L([x,y])=[L(x),L(y)] &  \\
L(x)(t(y))-L(y)(t(x))= t([x,y]) & \\
\end{cases}
\end{equation}
To see this, use the identification of $\al(x)$ with
$$\al(x)=\begin{pmatrix} L(x) & t(x) \\ 0 & 0 \end{pmatrix}.$$
Hence the Lie bracket in $\La \Lf \Lf (K^n)$ is given by
\begin{align*}
[\al(x),\al(y)] & =[(L(x),t(x)),(L(y),t(y))] \\
 & =(L(x)L(y)-L(y)L(x),L(x)(t(y))-L(y)(t(x)).
\end{align*}
It follows that $\al$ is a Lie algebra homomorphism if and only if $L$ is and $t$ is a
bijective $1$--cocycle in $Z^1(\Lg,\Lg_L)$.
The lemma follows with $t=\I$, the identity map on $\Lg$.
\end{proof}

What can we say about the existence of affine structures on Lie algebras ?

\begin{prop}\label{inv}
A finite-dimensional Lie algebra $\Lg$ admits an affine structure if and only if
there is a $\Lg$--module $M$ of dimension $\dim \Lg$ such that the vector
space $Z^1(\Lg,M)$ contains a nonsingular $1$--cocycle.
\end{prop}
\begin{proof}
Let $\phi \in Z^1(\Lg,M)$ be a nonsingular $1$-cocycle with inverse
transformation $\phi^{-1}$.
The module $M$ corresponds to a linear representation $\theta : \Lg \rightarrow
\Lg \Ll (\Lg)$. Then
\begin{equation*}
L(x):=\phi^{-1}\kringel \theta (x) \kringel \phi
\end{equation*}
defines a $\Lg$--module $N$ such that $\phi^{-1}\kringel\phi=\mathbf{1}
\in Z^1(\Lg,N)$.
It follows that $L:\Lg \rightarrow \Lg \Ll (\Lg)$ is a Lie algebra
representation and $\I([x,y])=\I(x)y-\I(y)x$
is a bijective $1$--cocycle in $Z^1(\Lg,\Lg_L)$. Hence $L(x)y=x\cdot y$
defines a left-symmetric structure on $\Lg$.
Conversely, $\I$ is a nonsingular $1$--cocycle if $\Lg$ admits a left-symmetric
structure.
\end{proof}

\begin{cor}
If the Lie algebra $\Lg$ admits a nonsingular
derivation, then there exists an affine structure on $\Lg$.
\end{cor}
\begin{proof}
Let $D$ be a nonsingular derivation and $\Lg$ the adjoint module of $\Lg$.
Since $Z^1(\Lg,\Lg)$ equals the space $\Der (\Lg)$ of derivations of $\Lg$,
$D$ is a nonsingular $1$-cocycle.
\end{proof}
\begin{cor}
If the Lie algebra $\Lg$ is graded by positive integers,
then there exists an affine structure on $\Lg$.
\end{cor}
\begin{proof}
Suppose that $\Lg=\oplus_{i\in \N}\; \Lg_i$ is a graduation, i.e.,
$[\Lg_i,\Lg_j]\subseteq \Lg_{i+j}$. Then there is a nonsingular
derivation defined by $D(x_i)=i x_i$ for $x_i\in \Lg_i$.
\end{proof}
\begin{cor}
Let $\Lg$ be a $2$-step nilpotent Lie algebra or a nilpotent Lie
algebra of dimension $n\le 6$. Then $\Lg$ admits an affine structure.
\end{cor}
\begin{proof}
It is well known that in both cases $\Lg$ can be graded by positive integers.
\end{proof}
The existence of a nonsingular derivation is a strong condition on the
Lie algebra. In fact, such a Lie algebra is necessarily
nilpotent \cite{JA1}. But not every nilpotent Lie algebra admits a
nonsingular derivation, see \cite{DIL}. The class of characteristically nilpotent
Lie algebras consists of nilpotent Lie algebras possessing only nilpotent
derivations. The example of a characteristically nilpotent Lie algebra,
given in \cite{DIL}, is $3$-step nilpotent.
Although there is no nonsingular derivation there exists an
affine structure. That follows from a theorem of Scheuneman \cite{SCH}:

\begin{prop}
Let $\Lg$ be a $3$-step nilpotent Lie algebra. Then
$\Lg$ admits an affine structure.
\end{prop}

For a new proof see \cite{BU3}. There have been attempts to generalize this result to $4$-step nilpotent
Lie algebras. However, only in special cases a positive result
could be proved, see \cite{DEH}, \cite{BU3}. The general case is still open. \\[0.3cm]
An affine structure on a Lie algebra implies the existence of a faithful
representation of relatively small degree:

\begin{prop}\label{faithful}
Let $\Lg$ be an $n$-dimensional Lie algebra over a field $K$ of
characteristic zero. If $\Lg$ admits an affine structure then $\Lg$
possesses a faithful Lie algebra module of dimension $n+1$.
\end{prop}

\begin{proof}
For any $\Lg$-module $V$ and any $\om \in Z^1(\Lg,V)$
we can define the $\Lg$-module $V_{\om}:=K\times V$ by the action
\begin{align*}
x\kringel (t,v) & = (0, x. v +t\om(x))
\end{align*}
where $x\in \Lg$, $t \in K$ and $v\in V$. It is easy to see that
\begin{align*}
x\kringel (y\kringel (t,v))- y\kringel (x\kringel (t,v)) & = [x,y]\kringel (t,v).
\end{align*}
We obtain a $\Lg$-module of dimension $\dim V+1$ which is faithful if
$\dim V=n$ and $\det \om \neq 0$. Hence if we just take $V=\Lg_L$ and
$\om =\I$, then $\I \in Z^1(\Lg,\Lg_L)$ because $\Lg$ admits a LSA-structure.
It follows that $V_{\om}$ is a faithful $\Lg$-module of dimension $n+1$.
\end{proof}

This proposition suggest to review Ado's Theorem, which states
that any finite-dimensional Lie algebra has a faithful finite-dimensional
representation:

\subsection{A refinement of Ado's theorem}

\begin{defi}
Let $\Lg$ be an $n$-dimensional Lie algebra over a field $K$ of
characteristic zero. Define an invariant of $\Lg$ by
$$\mu (\Lg,K):=\min \{\dim_K M \mid M \text{ is a faithful $\Lg$--module}\}.$$
\end{defi}

We write $\mu(\Lg)$ if the field is fixed.
By Ado's theorem, $\mu(\Lg)$ is finite. What can we say about the size of this
integer-valued invariant ? Following the details of the proof in Ado's theorem one obtains
an exponential bound on $\mu(\Lg)$, given by Reed \cite{REE}:

\begin{prop}
Let $\Lg$ be a solvable Lie algebra of dimension $n$ over an algebraically closed field 
of characteristic zero. Then $\mu (\Lg)\le n^n+n+1$.
\end{prop}
 
For semisimple Lie algebras we have  $\mu (\Lg)\le n$:
 
\begin{lem}
Let $\dim \Lg=n$. If $\Lg$ has trivial center then $\mu (\Lg)\le n$.
If $\Lg$ admits an affine structure then $\mu (\Lg)\le n+1$.
\end{lem}

\begin{proof}
The adjoint representation ${\rm ad}\colon \Lg\ra \Lg\Ll_n(K)$ is faithful if
and only if $\ker {\rm ad}=Z(\Lg)=0$. This yields a faithful $\Lg$-module of dimension
$n$. The second claim follows from proposition $\ref{faithful}$.
\end{proof}

For nilpotent Lie algebras the adjoint representation is not faithful. On the other hand
we know that all nilpotent Lie algebras $\Lg$ of class $2$ and $3$ admit an affine structure,
so that $\mu(\Lg)\le n+1$.
The following general bound for nilpotent Lie algebras has been given by Reed in
$1968$, see \cite{REE}:

\begin{prop}
Let $\Lg$ be a nilpotent Lie algebra of dimension $n$ and nilpotency class $k$. Then
$\mu (\Lg)\le n^k+1$.
\end{prop}

This bound is not very good. For filiform nilpotent Lie algebras we have $k=n-1$ and hence
$\mu (\Lg)\le n^{n-1}+1$. We have proved the following bound in $1997$, see \cite{BU7},
which is always better, for all $n\ge 2$ and all $2\le k \le n$:

\begin{thm}\label{pn1}
Let $\Lg$ be a nilpotent Lie algebra of dimension $n$ and nilpotency class
$k$. Denote by $p(j)$ the number of partitions of $j$ and let
$$p(n,k)=\sum_{j=0}^{k}\binom{n-j}{k-j}p(j).$$
Then $\mu(\Lg)\le p(n,k)$.
\end{thm}

Independently de Graaf \cite{GRA} proved the following bound, which is better than Reed's bound
but worse than ours:

\begin{thm}
Let $\Lg$ be a nilpotent Lie algebra of dimension $n$ and nilpotency class
$k$. Then $\mu(\Lg)\le \binom{n+k}{k}$.
\end{thm}

Fir fixed $k$, i.e., for  Lie algebras of constant nilpotency class $k$
these bounds are polynomial in $n$. For $k=1,\ldots , 5$ we have

\begin{align*}
p(n,1) & = n+1 \\[0.2cm]
p(n,2) & = \frac{1}{2}(n^2+ n+2)\\[0.2cm]
p(n,3) & = \frac{1}{6}(n^3+ 5n)\\[0.2cm]
p(n,4) & = \frac{1}{24}(n^4 - 2n^3+ 11n^2-10n+24)\\[0.2cm]
p(n,5) & =  \frac{1}{120}(n^5-5n^4+25n^3-55n^2+154n-240)
\end{align*}

\vspace*{0.5cm}
On the other hand we have, for $b(n,k)= \binom{n+k}{k}$,

\begin{align*}
b(n,1) & = n+1 \\[0.2cm]
b(n,2) & = \frac{1}{2}(n^2+ 3n +2)\\[0.2cm]
b(n,3) & = \frac{1}{6}(n^3+ 6n^2+11n+6)\\[0.2cm]
b(n,4) & = \frac{1}{24}(n^4 + 10n^3+ 35n^2+50n+24)\\[0.2cm]
b(n,5) & =  \frac{1}{120}(n^5+15n^4+85n^3+225n^2+274n+120)\\[0.2cm]
\end{align*}

Note that the $p(n,k)$ satisfy the following recursion
\[
p(n+1,k)=p(n,k)+p(n,k-1), \quad 1\le k\le n
\]
where we set $p(n,0)=1$. Indeed, we have

\begin{align*}
p(n,k)+p(n,k-1) & = \sum_{j=0}^{k}\binom{n-j}{k-j}p(j)+
\sum_{j=0}^{k-1}\binom{n-j}{k-j-1}p(j) \\
 & = \sum_{j=0}^{k-1}\left[\binom{n-j}{k-j}+\binom{n-j}{k-j-1}\right]p(j)+
\binom{n-k}{0}p(k)\\
 & = \sum_{j=0}^{k-1}\binom{n+1-j}{k-j}p(j) + p(k)\\
 & = p(n+1,k).
\end{align*}

The numbers $b(n,k)$ satisfy $b(n,1)<b(n,2)<\ldots < b(n,n)$. The behaviour of the
numbers $p(n,k)$ is quite different. We have proved the following in \cite{BU8}:

\begin{thm}\label{th2}
The function $p(n,k)$ is unimodal for fixed $n\ge 4$. More precisely we
have with $k(n)=\lfloor \frac{n+3}{2} \rfloor$
\begin{gather*}
p(n,1)<p(n,2)<\dots <p(n,k(n)-1)< p(n,k(n)),\\
p(n,k(n)) > p(n,k(n)+1)> \dots > p(n,n-1)>p(n,n).
\end{gather*}
\end{thm}

\begin{lem}
Let $F(q)=\prod_{j=1}^{\infty}(1-q^j)^{-1}$. For $2\le k\le n-1$ it holds
\begin{equation*}
p(n,k)<\binom{n}{k}F(\textstyle{\frac{k}{n}}).
\end{equation*}
\end{lem}

\begin{proof}
Denote by $p_k(j)$ the number of those partitions of $j$ in which each term
in the partition does not exceed $k$. We have
$$\sum_{j=0}^k p(j)q^j<\sum_{j=0}^{\infty}p_k(j)q^j=\prod_{j=1}^k\frac{1}
{1-q^j}$$ for $\abs q<1$. Hence
$$p(n,k)=\sum_{j=0}^k \binom{n-j}{k-j}p(j)<\sum_{j=0}^k
\binom{n}{k}q^jp(j)<\binom{n}{k}\prod_{j=1}^k\frac{1}{1-q^j}$$
with $q=k/n$.
\end{proof}

By estimating $p(n,k(n))$ we obtain the following result:

\begin{thm}\label{th3}
Let $\al=\frac{113}{40}$. Then
\begin{align*}
p(n,k) & < \frac{\al}{\sqrt{n}}\, 2^n \;\text{ for fixed $n\ge 1$ and all }
1\le k\le n.
\end{align*}
\end{thm}

If $k$ is depending on $n$, then the general bounds for $\mu(\Lg)$ 
are exponential in $n$. In this case it is harder to compare the bounds 
since we may have to consider how $k$ depends on $n$. 
For filiform Lie algebras this dependence is easy: $k=n-1$.
In that case our estimate for $\mu (\Lg)$ can be improved. In fact it holds
$\mu(\Lg)\le 1+p(n-2,n-2)$ which was the motivation to prove the
following propositions:

\begin{prop}\label{pr1}
Let $\al=\sqrt{2/3}\pi$. Then
$$p(n-1,n-1)<e^{\al \sqrt{n}}\quad \text{for all} \;\; n\ge 1.$$
\end{prop}

\begin{prop}\label{pr2}
Let $\al=\sqrt{2/3}\pi$. Then
$$p(n,n-1)<\sqrt{n}e^{\al \sqrt{n}} \quad \text{for all} \;\; n\ge 1.$$
\end{prop}

We obtain the following corollary:

\begin{cor}
Let $\Lg$ be a filiform nilpotent Lie algebra of dimension $n$ and  $\al=\sqrt{2/3}\pi$.
Then
\[
\mu(\Lg)<1+e^{\al\sqrt{n-1}}.
\]
\end{cor}

\begin{ex}
Let $\Lg={\rm span} \{x_1,\ldots ,x_6 \}$ with Lie brackets defined by
\[
[x_1,x_i]=x_{i+1}, \quad 2\le i\le 5
\]
\end{ex}
Then $\Lg$ is a $6$-dimensional Lie algebra of nilpotency class $5$.
For $n=4$ and $k=5$ the values of $n^k+1$,  $\binom{n+k}{k}$ and $p(n,k)$
are $7777$, $462$ and $45$ respectively. However the true size is known to
be $\mu(\Lg)=6$.\\[0.5cm]
In some cases we can determine $\mu(\Lg)$ by an explicit formula in the
dimension of $\Lg$. The first case is that
$\Lg$ is {\it abelian}. Then $\Lg$ is a vector space and any faithful
representation $\phi : \Lg \rightarrow \Lg\Ll(V)$, where $V$ is a
$d$--dimensional vector space, turns $\phi (\Lg)$ into an $n$--dimensional
commutative subalgebra of the matrix algebra $M_d(K)$.
There is an upper bound of $n$ in terms of $d$. Since $\phi$ is a
monomorphism, $n\le d^2$. A sharp bound was proved by Schur \cite{SHU}
over $\C$ and by Jacobson \cite{JA2} over any field $K$:
\begin{prop}\label{jac}
Let $M$ be a commutative subalgebra of $M_d(K)$ over an arbitrary field
$K$. Then $\dim M\le [d^2/4]+1$, where $[x]$ denotes the integral
part of $x$. This bound is sharp.
\end{prop}
Denote by $\lceil x \rceil$ the ceiling of $x$, i.e., the least integer
greater or equal than $x$.
\begin{prop}
Let $\Lg$ be an abelian Lie algebra of dimension $n$ over an arbitrary
field $K$. Then $\mu(\Lg)= \lceil 2\sqrt{n-1}\rceil $.
\end{prop}
\begin{proof}
By Proposition $\ref{jac}$, a faithful $\Lg$--module has dimension
$d$ with $n\le [d^2/4]+1$. This implies $d\ge \lceil 2\sqrt{n-1}\,\rceil $.
It is easy to construct commutative subalgebras of $M_d(K)$ of dimension
exactly equal to $[d^2/4]+1$. Hence $\mu(\Lg)=\lceil 2\sqrt{n-1}\,\rceil $.
\end{proof}

\begin{defi}
Let $\Lh_m(K)$ be a $(2m+1)$--dimensional vector space over $K$ with basis
$(x_1,\dots,x_m,y_1,\dots,y_m,z)$. Denote by $\Lh_m(K)$ the
$2$--step nilpotent Lie algebra defined by $[x_i,y_i]=z$ for $i=1,\dots,m$.
It is called {\it Heisenberg Lie algebra of dimension $2m+1$}.
\end{defi}

We have proved \cite{BU7}:

\begin{prop}
The Heisenberg Lie algebras satisfy $\mu (\Lh_m(K))=m+2$.
\end{prop}

\begin{prop}
Let $\Lg$ be a $2$--step nilpotent Lie algebra of dimension $n$ with
$1$--dimensional center. Then $n$ is odd and $\mu(\Lg)=(n+3)/2$.
\end{prop}
\begin{proof}
The commutator subalgebra $[\Lg,\Lg]\subseteq \Lz(\Lg)$ is $1$--dimensional.
Hence the Lie algebra structure on $\Lg$ is defined by a skew-symmetric
bilinear form $V\wedge V\rightarrow K$ where $V$ is the subspace of $\Lg$
complementary to $K=[\Lg,\Lg]$. It follows from the classification of such
forms that $\Lg$ is isomorphic to the Heisenberg Lie algebra $\Lh_m(K)$ with
$n=2m+1$. It follows $\mu(\Lg)=m+2=(n+3)/2$.
\end{proof}

Another important result about $\mu(\Lg)$ concerns the lower bounds for $\mu(\Lg)$.
Recall that any solvable Lie algebra $\Lg$ of dimension $n$
satisfying $\mu(\Lg)\ge n+2$ will be a counterexample to the Milnor
conjecture, because of proposition $\ref{faithful}$.
Unfortunately it turns out that it is non-trivial to find such Lie algebras.
It was known that filiform Lie algebras may be good candidates \cite{BEN}:

\begin{thm}
Let $\Lg$ be a filiform Lie algebra of dimension $n\ge 3$. Then 
$\mu(\Lg)\ge n$. 
\end{thm}

Studying these algebras in low dimensions yields \cite{BU6}: 

\begin{prop}
Let $\Lg$ be a filiform Lie algebra of dimension $n\le 9$.
Then $\mu(\Lg)= n$.
\end{prop}

Our main result in chapter $5$ of \cite{BU6} is:

\begin{thm}
There are families of filiform Lie algebras $\Lg$ of dimension $10\le n \le 13$
such that  $\mu(\Lg)\ge n+2$. Hence these Lie algebras do not admit any affine
structure.
\end{thm}

\subsection{The radical of an LSA}

If $G$ is a connected and simply connected Lie group acting simply transitively
as affine transformations on $\R^n$ then $G$ admits a complete left-invariant affine
structure. This means that the associated locally flat connection $\Delta$ on
$G$ is complete. As a consequence the Lie algebra $\Lg$ of $G$ is solvable \cite{AUS}
and the left-symmetric structure on $\Lg$ is complete.

\begin{defi}
The LSA $A$ is {\it complete} if for every $a\in A$ the linear transformation
$\I_A +R(a) \colon A \ra A$ is bijective.
\end{defi}

We have the following result, see \cite{SEG}:

\begin{thm}
Let $A$ be a finite-dimensional LSA over a field $K$ of characteristic zero.
Then the following conditions are equivalent:

\begin{itemize}
\item [(1)] $A$ is complete.
\item [(2)] $A$ is right nil, i.e., $R(x)$ is a nilpotent linear
transformation, for all $x \in A$.
\item [(3)] $R(x)$ has no eigenvalue in $K \setminus \{0\}$, for all $x \in A$.
\item [(4)] $\tr (R(x))=0$ for all $x \in A$.
\item [(5)] $Id + R(x)$ is bijective for all $x \in A$.
\end{itemize}
\end{thm}

The following definition is due to Koszul, see \cite{HEL}:

\begin{defi}
Let $A$ be an LSA and $T(A)=\{x\in A \mid \tr R(x)=0\}$. The largest left
ideal of $A$ contained in $T(A)$ is called the {\it radical} of $A$ and
is denoted by $\rad (A)$. 
\end{defi}

Note that $A$ is complete if and only if $A=\rad (A)$.
It is not clear whether this is a good definition of the radical of an
LSA. Usually the radical should be a $2$-sided ideal in the algebra.
Helmstetter \cite{HEL} has constructed an LSA $B$ where $\rad (B)$ is
not a $2$-sided ideal in general. Let $(A,\cdot)$ be an LSA and set 
\[
B=\End(A)\oplus A
\]
We may equipp this vector space with a left-symmetric product by
\begin{align*}
(f,a).(g,b) & = (fg+[L(a),g],a\cdot b+f(b)+g(a))
\end{align*}
for $a,b\in A$ and $f,g\in \End(A)$.

\begin{prop}
The algebra $B$ is an LSA. If $A$ is not complete then  $\rad (B)=0$.
If $A$ is complete and the product in $A$ is not identically zero then
$\rad(B)$ is not a $2$-sided ideal in $A$.
\end{prop}

However Mizuhara published results in \cite{MI1},\cite{MI2} claiming
that $\rad(A)$ is in fact a $2$-sided ideal in $A$ if the associated
Lie algebra $\Lg_A$ is solvable or nilpotent (over the complex numbers).
We have a counterexample for a $4$-dimensional LSA with solvable
Lie algebra.

\begin{ex}\label{notideal}
Define a $4$-dimensional left-symmetric algebra $A$ by the following product: 

\begin{center}
\bigskip\noindent
\begin{tabular}{ccccccccccc}
$e_1\cdot e_3$ & $=$ & $e_3$ & $\quad$ &  $e_2 \cdot e_2$ & $=$ & $2e_2$  & $\quad$ &  $e_3 \cdot e_4$ & $=$ & $e_2$ \\ 
$e_1\cdot e_4$ & $=$ & $-e_4$ & $\quad$ &  $e_2 \cdot e_3$ & $=$ & $e_3$  & $\quad$ &  $e_4 \cdot e_3$ & $=$ & $e_2$ \\
               &     &        & $\quad$ &  $e_2 \cdot e_4$ & $=$ & $e_4$  & $\quad$ &   &  & \\[0.3cm]  
\end{tabular}
\end{center}
and the other products equal to zero. 
Then $\rad (A)=\s \{ e_1\}$. This is not a right 
ideal in $A$.
\end{ex}

The right multiplications are given by

\begin{align*}
R(e_1)& =\begin{pmatrix} 0 & 0 & 0 & 0 \\
                       0 & 0 & 0 & 0 \\
                       0 & 0 & 0 & 0 \\    
                       0 & 0 & 0 & 0 
\end{pmatrix},\quad
R(e_1)=\begin{pmatrix} 0 & 0 & 0 & 0 \\
                       0 & 2 & 0 & 0 \\
                       0 & 0 & 0 & 0 \\    
                       0 & 0 & 0 & 0 
\end{pmatrix}, \\[0.5cm]
R(e_3)& =\begin{pmatrix} 0 & 0 & 0 & 0 \\
                       0 & 0 & 0 & 1 \\
                       1 & 1 & 0 & 0 \\    
                       0 & 0 & 0 & 0 
\end{pmatrix},\quad
R(e_4)=\begin{pmatrix} 0 & 0 & 0 & 0 \\
                       0 & 0 & 1 & 0 \\
                       0 & 0 & 0 & 0 \\    
                       -1 & 1 & 0 & 0 
\end{pmatrix}. \\
\end{align*}

We see that $T(A)=\ker \tr R= \s \{ e_1,e_3,e_4 \}$. The largest left ideal in $T(A)$ is given by $\s \{ e_1\}$.
The solvable, non-nilpotent Lie algebra is given by
\begin{align*}
[e_1,e_3]=e_3, \quad  [e_2,e_3]= e_3,\\
[e_1,e_4]=-e_4, \quad   [e_2,e_4]= e_4.\\
\end{align*}

\begin{rem}
The above counterexample can be generalized to all dimensions $n\ge 4$.
We do not know of a counterexample for an LSA $A$ if
the Lie algebra $\Lg_A$ is nilpotent. In \cite{MI1} it is claimed that
$\rad(A)$ is a $2$-sided ideal in $A$ containing $[A,A]$ if $\Lg_A$ is
nilpotent over the real numbers. 
\end{rem}

There are several other possibilities for radicals of an LSA.

\begin{defi}
Let $A$ be an arbitrary finite-dimensional algebra and $I$ an ideal in $A$.
Define sets $I^{(k)}$ inductively by $I^{(0)}=I$ and  $I^{(i+1)}= I^{(i)} I^{(i)}$.
Denote by $^kI$ the linear span of all elements $L(a_1)L(a_2)\cdots L(a_{k-1})a_k$ for all
$a_1,\ldots,a_k \in I$. An ideal $I$ is called {\it solvable}, if  $I^{(k)}=0$ for some
$k\ge 0$. It is called {\it left-nilpotent} if $^kI=0$ for some $k\ge 1$. 
\end{defi}

Note that any left-nilpotent ideal is solvable.
If $I$ and $J$ are solvable ideals in $A$ then $I+J$ is again a solvable ideal in $A$.
Hence there exists a unique maximal solvable ideal of $A$. In particular the following
definition makes sense.

\begin{defi}
Let $A$ be a finite-dimensional LSA. Then the {\it solvable radical} $\sol(A)$ of $A$ is
the unique maximal solvable ideal of $A$.
\end{defi}

Unlike the solvable case there is in general no guarantee for the existence of a unique maximal
left-nilpotent ideal in $A$. For LSAs however we have the following result \cite{CKL}:

\begin{lem}
Let $A$ be a finite-dimensional LSA. If $I$ and $J$ are left-nilpotent ideals of $A$,
then so is $I+J$.
\end{lem}

\begin{cor}
If $A$ is a finite-dimensional LSA, then $A$ has a unique maximal left-nilpotent
ideal $\nil(A)$ containing all left-nilpotent ideals of $A$. It is called the
left-nilpotent radical of $A$ and satisfies $\nil(A)\subseteq \sol(A)$.
\end{cor}

The last claim follows from the fact that left-nilpotent ideals in $A$ are sol\-vable.
Let us consider now the symmetric bilinear form $s$ on $A$ defined by
\[
s(x,y)=\tr R(x)R(y).
\]
Its radical is given by
\[  
A^{\perp}=\{ a\in A\mid s(a,b)=0 \; \; \forall \, b\in A \}.
\]
Unfortunately, this need not be an ideal for an LSA $A$. Also it need not
coincide with the Koszul radical $\rad(A)$ of $A$.
But we have the following result \cite{CKL}:

\begin{thm}
Let $A$ be a finite-dimensional LSA over $\R$. Then we have the relations
\[
\nil(A)\subseteq \rad(A)\subseteq A^{\perp} \subseteq T(A).
\]
\end{thm}

\begin{cor}
The LSAs $\nil(A), \rad(A)$ and  $A^{\perp}$ are complete, and $\rad(A)$
is the maximal complete left ideal of $A$.
\end{cor}

\begin{ex}
Let $A$ be the LSA of example \eqref{notideal}. Then $\nil(A)=0$,
$\rad(A)=A^{\perp}=\s \{ e_1\}$ and $T(A)=\s \{ e_1,e_3,e_4 \}$.
\end{ex}

Indeed, since $\rad(A)$ is $1$-dimensional and not an ideal in $A$,
the ideal $\nil(A)$ must be zero. The different radicals of $A$ need not
be equal in general. However the following result is known \cite{KIM}, \cite{HEL}:

\begin{lem}
Let $A$ be a finite-dimensional LSA over $\R$. Then the following conditions
are equivalent:
\begin{itemize}
\item [(1)] $A$ is left-nilpotent.
\item [(2)] $A$ is complete and $\Lg_A$ is nilpotent.
\item [(3)] $L(x)$ is a nilpotent transformation, for all $x\in A$.
\end{itemize}
\end{lem}

Suppose that the Lie algebra $\Lg_A$ is nilpotent. Since $\rad(A)$ is complete
the Lemma implies that $\rad(A)$ is left-nilpotent. 
If we believe that $\rad(A)$ is an ideal in this case, it follows that
$\rad(A)\subseteq \nil(A)$ and hence $\rad(A)= \nil(A)$.
More generally the following result is proved in \cite{CKL2}:

\begin{thm}
Let $A$ be a finite-dimensional LSA over $\R$ or $\C$. Let
$S=\{ a\in A \mid R(a)$ is nil\-po\-tent $\}$.
If $\Lg_A$ is nilpotent then
\[ 
\nil(A)=\rad(A)=A^{\perp}=S.
\]
\end{thm}

\subsection{Simple LSAs}

Let $A$ be an LSA over $K$ of dimension $n\ge 2$ and assume that the product is non-trivial.
Denote by $\Lg_A$ the Lie algebra of $A$.

\begin{defi}
The algebra $A$ is called {\it simple} if every two-sided ideal in $A$ is equal to
$A$ or equal to $0$.
\end{defi}

Recall that the map $L \colon  \Lg \rightarrow \Lg \Ll (A)$ with
$x \mapsto L(x)$ is a Lie algebra representation, i.e.,
$ L([x,y]) = [L(x),L(y)] $. It is easy to see that $\ker (L)$ is a two-sided
ideal in $A$. If $A$ is simple then $\ker(L)=0$, since we assume that the product of
$A$ is non-trivial. This yields the following result.

\begin{lem}
Let $A$ be a simple, non-trivial LSA of dimension $n$. Then we have $\mu(\Lg_A)\le n$ 
for the Lie algebra $\Lg_A$ of $A$.
\end{lem}

Indeed, the left multiplication $L$ is a faithful representation of dimension $n$ since
$\ker(L)$ is zero.\\
Of course we have many examples of simple LSAs. Just consider simple associative algebras.
The following lemma yields also different examples of simple LSAs:

\begin{lem}
Let $A$ be an LSA with reductive Lie algebra of $1$-dimensional center. Then $A$ is simple.
\end{lem}

\begin{proof}
Let $\Lg=\Lg_A=\Ls \oplus \Lz$ be the Lie algebra with center $\Lz \simeq K$.
Suppose $I$ is a proper two--sided ideal in $A$. Then it is also
a proper Lie ideal in $\Lg$ and both $I$ and $\Lg/I$ inherit an LSA--structure
from $A$. Since a semisimple Lie algebra does not admit any LSA-structures,
we conclude that $I$ must be equal to $\Ls_1 \oplus K$, where $\Ls_1$ is a
semisimple ideal of $\Ls$. Hence $\Lg/I$ is semisimple and admits an
LSA--structure. This is a contradiction. 
\end{proof}

Now there exist infinitely many non-isomorphic LSA-structures on 
$\Lg\Ll(n,K)$, which have been classified in \cite{BU9}, \cite{BAU}. They are simple
as LSAs, not necessarily associative, and they all arise by deformations of the associative matrix algebra structure.
The question is whether all simple simple LSAs must have a reductive Lie algebra.
This is not the case:

\begin{ex}
Define an $n$-dimensional LSA $A$ with basis $(e_1,\ldots e_n)$ by the following product: \\
\begin{center}
\begin{tabular}{ccccccc}
$e_1\cdot e_1$ & $=$ & $2e_1$ & $\quad$ &  $e_j \cdot e_j$ & $=$ & $e_1$, $\quad j=2,\ldots ,n  $\\ 
$e_1\cdot e_j$ & $=$ & $e_j$ &  $\quad$ &   &  &  \\[0.3cm]
\end{tabular}
\end{center}
and the other products equal to zero. Then $A$ is a simple, incomplete LSA with two-step solvable
Lie algebra.
\end{ex}

Let $I$ be a non-zero ideal in $A$ and $x\in I$. Then $e_j\cdot x$ is a multiple of $e_1$ for
each $j\ge 2$. It follows that $e_1\in I$ and hence $I=A$. Hence $A$ is simple. It is not complete since
$\tr R(e_1)=2$. The Lie algebra $\Lg_A$ is two-step solvable with brackets $[e_1,e_j]=e_j$ for $j\ge 2$.\\[0.3cm]
What can we say on the Lie algebra of a simple LSA ?

\begin{lem}
Let $A$ be an LSA with Lie algebra $\Lg_A$. Then $\Lg_A$ is abelian if and only if $A$ is associative
and commutative.
\end{lem}

\begin{proof}
If $A$ is commutative then $\Lg$ is abelian by definition.
Assume that $\Lg$ is abelian. Then $x.y=y.x$ for all
$x,y \in A$ and using left--symmetry, $0=[xz].y=x.(z.y)-z.(x.y)=
x.(y.z)-(x.y).z=(x,y,z)$.
\end{proof}

In particular the Lie algebra of a simple LSA cannot be abelian since $A$ is not
one-dimensional. This result can be generalized as follows \cite{BU2}:

\begin{prop}
If $A$ is a simple LSA then $\Lg_A$ cannot be nilpotent.
\end{prop}

The classification of simple LSAs is only known in low dimensions. Up to LSA-isomorphism
there is only one 2-dimensional simple complex LSA. It is given by $A=\C x\oplus \C y$ with product 
\[
x.x=2x,\; x.y=y,\; y.x=0,\; y.y=x.
\]
In dimension $3$ the classification is as follows, see \cite{BU2}:

\begin{prop}
Let $A$ be a simple $3$-dimensional LSA over $\C$. Then its Lie algebra $\Lg$ is isomorphic to
$\Lr_{3,\la}=<e_1,e_2,e_3 \mid [e_1,e_2]=e_2,\,[e_1,e_3]=\la e_3>$ with $|\la| \le 1,\la \ne 0$, 
and $A$ is isomorphic to exactly one of the following algebras $A_{1,\la}$ and $A_2$:

\begin{center}
\bigskip\noindent
\begin{tabular}{ccccccccccc}
$e_1\cdot e_1$ & $=$ & $(\la+1)e_1$ & $\quad$ &  $e_1 \cdot e_3$ & $=$ & $\la e_3$ & $\quad$ & $e_3 \cdot e_2$ 
& $=$ & $e_1$ \\ 
$e_1\cdot e_2$ & $=$ & $e_2$ & $\quad$ &  $e_2 \cdot e_3$ & $=$ & $e_1$  & $\quad$ &  & & \\[0.3cm]
\end{tabular}
\end{center}

\begin{center}
and
\end{center}

\begin{center}
\bigskip\noindent
\begin{tabular}{ccccccccccc}
$e_1\cdot e_1$ & $=$ & $\frac{3}{2}e_1$ & $\quad$ &  $e_1 \cdot e_3$ & $=$ & $\frac{1}{2}e_3$ & $\quad$ & $e_3 \cdot e_2$ 
& $=$ & $e_1$ \\ 
$e_1\cdot e_2$ & $=$ & $e_2$ & $\quad$ &  $e_2 \cdot e_3$ & $=$ & $e_1$  & $\quad$ & $e_3\cdot e_3$ & $=$ & 
$-e_2$ \\[0.3cm]
\end{tabular}
\end{center}
\end{prop}

\begin{cor}
Let $A$ be a complete simple LSA of dimension $3$ over $\C$. Then $A$ is isomorphic to $A_{1,-1}$ with
Lie algebra $\Lr_{3,-1}(\C)$.
\end{cor}

The classification of simple LSAs in dimension $4$ is quite complicated. It is much
easier to consider the complete ones here: any $4$-dimensional complete simple LSA over $\C$ is isomorphic
to the following LSA, see \cite{BU2}:
\begin{center}
\bigskip\noindent
\begin{tabular}{ccccccccccc}
$e_1\cdot e_2$ & $=$ & $e_4$ & $\quad$ &  $e_3 \cdot e_2$ & $=$ & $e_1$  & $\quad$ &  $e_4 \cdot e_3$ & $=$ & $2e_3$ \\ 
$e_2\cdot e_1$ & $=$ & $e_4$ & $\quad$ &  $e_4 \cdot e_1$ & $=$ & $e_1$  & $\quad$ &   &  &  \\
$e_2\cdot e_3$ & $=$ & $e_4$ & $\quad$ &  $e_4 \cdot e_2$ & $=$ & $-e_2$  & $\quad$ &   &  & \\[0.3cm]  
\end{tabular}
\end{center}

It is possible to associate certain weights and graphs for so called ``special'' complete LSAs, see \cite{BU2}.
The above algebra has weights $\Lambda=\{-1,0,1,2 \}$, and the graph is given by 

\begin{center}
\scalebox{0.7}{\includegraphics{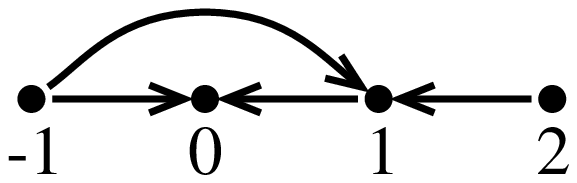}}
\end{center}

This gives some idea how to classify special complete simple LSAs in general.

\subsection{Cohomology of LSAs}

Let $A$ be an LSA and denote  by $C^n(A,A)= \{ f \,:\, A \times \cdots \times A \rightarrow A \mid f 
\hbox{ is multilinear} \}$ be the space of $n$-cochains, where $A$ is the regular module for $A$. Define the
coboundary operator $\de^n \, : \, C^n(A,A) \rightarrow C^{n+1}(A,A)$ by

\begin{align*}
(\de^nf)(x_1,\ldots,x_{n+1}) & =\sum_{i=1}^n (-1)^{i+1}x_i.f(x_1,\ldots,\widehat{x_i},\ldots ,x_{n+1}) \\
 & + \sum_{i=1}^n (-1)^{i+1}f(x_1,\ldots,\widehat{x_i},\ldots ,x_n,x_i).x_{n+1} \\
 & - \sum_{i=1}^n (-1)^{i+1}f(x_1,\ldots,\widehat{x_i},\ldots ,x_n,x_i.x_{n+1}) \\
 & + \sum_{i<j\le n} (-1)^{i+j}f([x_i,x_j],x_2,\ldots ,\widehat{x_i},\ldots , \widehat{x_j},\ldots ,x_{n+1}).\\
\end{align*} 

In particular we have

\begin{align*}
(\de^1f)(x_1,x_2) & = x_1.f(x_2)+f(x_1).x_2-f(x_1.x_2) \\
(\de^2f)(x_1,x_2,x_3) & = f(x_1,x_2.x_3)-f(x_1.x_2,x_3)+f(x_2.x_1,x_3)- f(x_2,x_1.x_3) \\
 & + x_1.f(x_2,x_3) - f(x_1,x_2).x_3+f(x_2,x_1).x_3 -x_2.f(x_1,x_3).\\
\end{align*} 
Recall that $[x_i,x_j]=x_i.x_j-x_j.x_i$. Since $\de^2=0$ we obtain cohomology groups $H^n_{LSA}(A,A)$.
Note that $Z^1(A,A)=\Der (A)$, and that $Z^2(A,A)$ describes infinitesimal left-symmetric deformations of $A$,
in the sense of Gerstenhaber. Nijenhuis showed in \cite{NI2}, that many properties of this LSA-cohomology
can be deduced from Lie algebra cohomology. In fact, we have
\begin{align*}
H^n_{LSA}(A,A) & \cong H^{n-1}(\Lg_{A}, \End(A)),
\end{align*}
where $\Lg_A$ denotes the underlying Lie algebra of $A$. Dzhumadil'daev \cite{DZ2} more generally has defined
cohomology groups $H^n_{RSA}(A,M)$ for arbitrary right-symmetric modules $M$. He proves, among other things that
\begin{align*}
H^n_{RSA}(A,M) & \cong H^{n-1}(\Lg_{A}, C^1(A,M)).
\end{align*}

\begin{ex}
Let $A$ be an RSA with Lie algebra $\Lg_A=\Lg\Ll_n(K)$ over a field $K$ of characteristic zero
(see \cite{BAU}, \cite{BU9}, \cite{DZ2}). Then, for $k\ge 1$, \\
\begin{align*}
Z^1_{RSA}(A,A) & \cong Z^1(\Ls\Ll_n(K),\Ls\Ll_n(K))\cong \Ls\Ll_n(K),\\
H^k_{RSA}(A,A) & \cong Z^1_{RSA}(A,A) \otimes H^{k-1}(\Lg\Ll_n(K),K).
\end{align*}
\end{ex}


\begin{thebibliography}{99}    


\bibitem{AUS} L. Auslander: {\it Simply transitive groups of affine
motions}. Am.\ J.\ of Math.\ \textbf{99} (1977), 809-826.

\bibitem{BAM} C. Bai, D. Meng, {\it A Lie algebraic approach to Novikov algebras}. 
J.\ Geom.\ Phys.\textbf{45} (2003), no. 1-2, 218-230. 

\bibitem{BAN} A. A. Balinskii; S. P. Novikov: {\it Poisson brackets of
hydrodynamic type, Frobenius algebras and Lie algebras}.
Sov.\ Math.\ Dokl.\ \textbf{32} (1985), 228-231.

\bibitem{BBK} B. Bakalov; V. Kac: {\it Field algebras}.
Int.\ Math.\ Res.\ Not. \textbf{2003}, no. 3, 123-159.

\bibitem{BAU} O. Baues: {\it Left-symmetric algebras for $\Lg\Ll(n)$}. 
 Trans.\ Amer.\ Math.\ Soc.\  \textbf{351} (1999), no.7, 2979-2996. 

\bibitem{BEN} Y. Benoist, {\it Une nilvari\'et\'e non affine}. J.\ Differential
Geom. \textbf{41} (1995), 21-52.

\bibitem{BEZ} J. P. Benz\'ecri: {\it Vari\'et\'es  localement affines}.
Th\`ese, Princeton Univ.\ , Princeton, N.\ J.\ (1955).

\bibitem{BOR} R. E. Borcherds: {\it Vertex algebras, Kac-Moody algebras, and the Monster}. 
Proc.\ Nat.\ Acad.\ Sci.\ \textbf{83} (1986), no. 10, 3068-3071.

\bibitem{BOY} N. Boyom: {\it Sur les structures affines homotopes
\`a z\'ero des groupes de Lie}. J.\ Diff.\ Geom.\ \textbf{31} (1990), 859-911.

\bibitem{BU1} D. Burde: {\it Affine structures on nilmanifolds}. Int.\ J.\
of Math.\ \textbf{7} (1996), 599-616.

\bibitem{BU2} D. Burde: {\it Simple left-symmetric algebras with solvable Lie
algebra}. Manuscripta Mathematica {\bf 95} (1998), 397-411.

\bibitem{BU3} D. Burde, K. Dekimpe: {\it Novikov structures on solvable Lie algebras}.  
to appear in J.\ Geom.\ Phys.\ (2006).

\bibitem{BU4} D. Burde, F. Grunewald: {\it Modules for certain Lie algebras of maximal class}.
J.\ Pure Appl.\ Algebra \textbf{99} (1995), 239-254.

\bibitem{BU5} D. Burde: {\it Affine cohomology classes for filiform
Lie algebras}. Contemporary Mathematics \textbf{262} (2000), 159-170.

\bibitem{BU6} D. Burde: {\it Left-invariant affine structures on nilpotent Lie groups}. 
Habilitation thesis, D\"usseldorf (1999).

\bibitem{BU7} D. Burde: {\it A refinement of Ado's Theorem}. Archiv Math.\
\textbf{70} (1998), 118-127.

\bibitem{BU8} D. Burde: {\it Estimates on binomial sums of partition functions}. Manuscripta mathematica 
\textbf{103} (2000), 435-446.

\bibitem{BU9} D. Burde: {\it Left-invariant affine structures on reductive Lie groups}. J. Algebra 
\textbf{181} (1996),884-902.

\bibitem{CAY} A. Cayley: {\it On the Theory of Analytic Forms Called Trees}.
Collected Mathematical Papers of Arthur Cayley, Cambridge Univ. Press, Cambridge, 1890, Vol. 
\textbf{3} (1890), 242-246.

\bibitem{CDM} Y. Carri\'ere, F. Dal'bo, G. Meigniez: {\it Inexistence de
structures affines sur les fibres de Seifert}. Math.\ Ann.\ \textbf{296} (1993), 743-753.

\bibitem{CKL} K. S. Chang, H. Kim, H. Lee: {\it On radicals of a left-symmetric algebra}.
Commun.\ Algebra \textbf{27} (1999), no.7, 3161-3175.

\bibitem{CKL2} K. S. Chang, H. Kim, H. Lee: {\it Radicals of a left-symmetric algebra on a nilpotent
Lie group}. Bull.\ Korean Math.\ Soc.\  \textbf{41} (2004), no.2, 359-369.

\bibitem{CHL} F. Chapoton, M. Livernet: {\it Pre-Lie algebras and the rooted trees operad}. 
 Intern.\ Math.\ Research Notices \textbf{8} (2001), 395-408.

\bibitem{COK} A. Connes, D. Kreimer: {\it Hopf algebras, renormalization 
and noncommutative geometry}. Comm.\ Math.\ Phys.\  \textbf{199} (1998), no. 1, 203-242.

\bibitem{DEH} K. Dekimpe, M. Hartl: {\it Affine structures on $4$--step
nilpotent Lie algebras}. J.\ Pure Appl.\ Math.\ \textbf{129} (1998), 123-134.

\bibitem{DEM} K. Dekimpe, W. Malfait: {\it Affine structures on a class of
virtually nilpotent groups}. Topology Appl.\ \textbf{73} (1996), 97-119.

\bibitem{DEK} A. De Sole, V. G. Kac: {\it Freely generated vertex algebras and non-linear Lie conformal
algebras}. Comm.\ Math.\ Phys.\  \textbf{254}  (2005),  no. 3, 659-694. 

\bibitem{DIL} J. Dixmier, W. G. Lister: {\it Derivations of nilpotent
Lie algebras}. Proc.\ Amer.\ Math.\ Soc.\ \textbf{8} (1957), 155-158.

\bibitem{DOR} J. Dorfmeister: {\it Quasi-clans}. 
Abh.\ Math.\ Semin.\ Univ.\ Hamburg \textbf{50} (1980), 178-187.

\bibitem{DZL} A. Dzhumaldil'daev, C. Löfwall: {\it Trees, free right-symmetric algebras, free 
Novikov algebras and identities}.
Homology Homotopy Appl.\ \textbf{4} (2002), no. 2(1), 165-190.

\bibitem{DZ1} A. Dzhumaldil'daev: {\it $N$-commutators}. 
Comment.\ Math.\ Helv.\ \textbf{79} (2004), no. 3, 516-553. 

\bibitem{DZ2} A. Dzhumaldil'daev: {\it Cohomologies and deformations of right-symmetric algebras}. 
J.\ Math.\ Sci.\textbf{93} (1999),  no. 6, 836-876.

\bibitem{FHL} I. B. Frenkel, Y. Huang, J. Lepowsky: {\it On axiomatic approaches to 
vertex operator algebras and modules}. Mem.\ Amer.\ Math.\ Soc.\textbf{104} (1993), 
no. 494, 1-64. 

\bibitem{FLM} I. B. Frenkel, J. Lepowsky, A. Meurman: {\it Vertex operator algebras and the Monster}. 
Pure and Applied Mathematics, \textbf{134} (1988), Academic Press, Boston, MA, 1-508.

\bibitem{GER} M. Gerstenhaber: {\it The cohomology structure of an associative ring}. 
Ann.\ of Math.\ \textbf{78} (1963), 267-288.

\bibitem{GRA} W. A. de Graaf: {\it Constructing faithful matrix representations of Lie algebras}.
 Proceedings of the 1997 International Symposium on Symbolic and Algebraic Computation,  
54-59 (electronic), ACM, New York.

\bibitem{HEL} J. Helmstetter: {\it Radical d'une alg\`ebre sym\'etrique
a gauche}. Ann.\ Inst.\ Fourier \textbf{29} (1979), 17-35.

\bibitem{JA1} N. Jacobson: {\it A note on automorphisms and derivations
of Lie algebras}. Proc.\ Amer.\ Math.\ Soc.\ \textbf{6} (1955), 281-283.

\bibitem{JA2} N. Jacobson: {\it Schur's theorem on commutative matrices}.
Bull.\ Amer.\ Math.\ Soc.\ \textbf{50} (1944), 431-436.

\bibitem{KAC} V. Kac: {\it Vertex algebras for beginners}. University Lecture Series \textbf{10}. 
American Mathematical Society, Providence (1998), 1-201.

\bibitem{KIM} H. Kim: {\it Complete left-invariant affine structures
on nilpotent Lie groups}. J.\ Diff.\ Geom.\ \textbf{24} (1986), 373-394.

\bibitem{KOB} S. Kobayashi, K. Nomizu: {\it Foundations of Differential
Geometry}. Vols. I and II, Wiley-Interscience Publishers, New York and
London (1969).

\bibitem{KOS} J-L. Koszul: {\it Domaines born\'es homog\`enes et orbites de groupes de transformations 
affines}. Bulletin de la Soci\'et\'e Math\'ematique de France \textbf{89} (1961), p. 515-533 

\bibitem{KR1} D. Kreimer: {\it New mathematical structures in renormalizable quantum 
field theories}.  Ann.\ Physics  \textbf{303} (2003), no. 1, 179-202. 

\bibitem{KR2} D. Kreimer: {\it Structures in Feynman Graphs - Hopf Algebras and Symmetries}.  
Proc.\ Symp..\ Pure Math.\ \textbf{73} (2005), 43-78. 

\bibitem{KUI} N. H. Kuiper: {\it Sur les surfaces localement affines}.
Colloque de G\'eometrie diff\'erentielle, Strasbourg (1953), 79-86.

\bibitem{LEL} J. Lepowsky, H. Li: {\it Introduction to Vertex Operator Algebras and Their 
Representations}. Progress in Mathematics Vol. \textbf{227}, Birkh\"auser (2003), 1-316.

\bibitem{MAY} J. P. May: {\it Geometry of Iterated Moduli Spaces}.
Lecture Notes in Math.\  \textbf{271} (1972). 

\bibitem{MIL} J. Milnor: {\it On fundamental groups of complete affinely
flat manifolds}. Advances in Math.\ \textbf{25} (1977), 178-187.

\bibitem{MI1} A. Mizuhara: {\it On the radical of a left-symmetric
algebra}. Tensor N.\ S.\ \textbf{36} (1982), 300-302.

\bibitem{MI2} A. Mizuhara: {\it On the radical of a left-symmetric
algebra II}. Tensor N.\ S.\ \textbf{40} (1983), 221-232.

\bibitem{NAY} T. Nagano, K. Yagi: {\it The affine structures on the real
two torus}. Osaka J.\ Math.\ \textbf{11} (1974), 181-210.

\bibitem{NI1} A. Nijenhuis: {\it The graded Lie algebras of an algebra.} 
Indag.\ Math.\ \textbf{29} (1967), 475-486.

\bibitem{NI2} A. Nijenhuis: {\it On a class of common properties of some different types of algebras. II.} 
Nieuw Arch.\ Wisk.\ (3)  \textbf{17}  (1969), 87-108. 

\bibitem{NIS} M. Nisse: {\it Structure affine des infranilvari\'et\'es et
infrasolvari\'et\'es}. C.\ R.\ Acad.\ Sci.\ Paris \textbf{310} (1990), 667-670.

\bibitem{OS1} J. M. Osborn: {\it Novikov algebras}.
Nova J.\ Algebra Geom.\ \textbf{1} (1992), no. 1, 1-13. 

\bibitem{OS2} J. M. Osborn: {\it Infinite dimensional Novikov algebras of characteristic $0$}.
J.\ Algebra \textbf{167} (1994), no. 1, 146-167.

\bibitem{REE} B. E. Reed: {\it Representations of solvable Lie algebras}.
Michigan Math.\ J.\ \textbf{16} (1969), 227-233.

\bibitem{ROS} M. Rosellen: {\it A course in vertex algebra}. Preprint (2005).

\bibitem{SCH} J. Scheuneman: {\it Affine structures on three-step
nilpotent Lie algebras}. Proc.\ Amer.\ Math.\ Soc.\ \textbf{46} (1974), 451-454.

\bibitem{SHU} I. Schur: {\it Zur Theorie vertauschbarer Matrizen}.
J.\ Reine Angew.\ Mathematik \textbf{130} (1905), 66-76.

\bibitem{SEG} D. Segal: {\it The structure of complete left-symmetric
algebras}. Math.\ Ann.\  \textbf{293} (1992), 569-578.

\bibitem{SMI} J. Smillie: {\it An obstruction to the existence of affine
structures}. Invent.\ Math.\ \textbf{64} (1981), 411-415.

\bibitem{THU} W. P. Thurston: {\it Three-dimensional Geometry and
Topology Vol. 1}. Princeton Mathematical Series \textbf{35},
Princeton University Press (1997).

\bibitem{VIN} E. B. Vinberg: {\it Convex homogeneous cones}. Transl.\ Moscow
Math.\ Soc.\ \textbf{12} (1963), 340-403.

\bibitem{ZEL} E. Zelmanov: {\it  On a class of local translation invariant
Lie algebras}. Soviet Math. Dokl. {\bf 35} (1987), 216-218.



\end{thebibliography}
\end{document}